\documentclass[onecolumn]{aa}
\usepackage[varg]{txfonts}

\newcommand{\crra}{Cram\'er-Rao}

\newcommand{\R}{\mathbb{R}}

\newcommand{\argmin}{\arg\!\min}
\usepackage{epstopdf}
\usepackage{xcolor}

\usepackage{amsmath, amssymb, dsfont}

\begin{document}
\title{Optimality of the Maximum Likelihood estimator in Astrometry}

\author{Sebastian Espinosa\inst{1} \and Jorge F. Silva\inst{1} \and
  Rene A. Mendez\inst{2} \and Rodrigo
  Lobos\inst{3} \and Marcos Orchard\inst{1} }

\institute{Information and Decision Systems Group, Department of
  Electrical Engineering, Facultad de Ciencias F\'{\i}sicas y
  Matem\'aticas, Universidad de Chile, Beauchef 850, Santiago, Chile \\
\email{sebastian.espinosa, josilva, morchard @ing.uchile.cl}
\and Departamento de Astronom\'{\i}a, Facultad de Ciencias F\'{\i}sicas y
Matem\'aticas, Universidad de Chile, Casilla 36-D, Santiago, Chile \\
\email{rmendez@uchile.cl} \and Department of Electrical Engineering,
University of Southern California, 90007, California, USA \\
\email{rlobos@usc.edu}}

\abstract{Astrometry relies on the precise measurement of positions
  and motions of celestial objects. Driven by the ever-increasing
  accuracy of astrometric measurements, it is
  important to critically assess the maximum precision that could be
  achieved with these observations.}
{The problem of astrometry is revisited from the perspective of
  analyzing the attainability of well-known performance limits (the
  \crra\ bound) for the estimation of the relative position of
  light-emitting (usually point-like) sources on a CCD-like detector
  using commonly adopted estimators such as the weighted
  least squares and the maximum likelihood.}
{Novel technical results are presented to determine the performance of
  an estimator that corresponds to the solution of an optimization
  problem in the context of astrometry. Using these results we are
  able to place stringent bounds on the bias and the variance of
  the estimators in close form as a function of the data. We confirm
  these results through comparisons to numerical simulations under a
  broad range of realistic observing conditions.}
{The maximum likelihood and the weighted least
  square estimators are analyzed. We confirm the sub-optimality of the
  weighted least squares scheme from medium to high signal-to-noise
  found in an earlier study for the (unweighted) least
  squares method. We find that the maximum likelihood estimator
  achieves optimal performance limits across a wide range of relevant
  observational conditions. Furthermore, from our results, we provide concrete insights for adopting an adaptive
  weighted least square estimator that can be regarded as a
  computationally efficient alternative to the optimal maximum
  likelihood solution.}
{We provide, for the first time, close-form analytical expressions
  that bound the bias and the variance of the weighted least square
  and maximum likelihood implicit estimators for astrometry using a
  Poisson-driven detector. These expressions can be used to formally
  assess the precision attainable by these estimators in comparison
  with the minimum variance bound.}

\keywords{Astrometry, parameter estimation, information limits,  \crra\ bound,
maximum likelihood, weighted least squares.}

\maketitle

\section{Introduction}
\label{sec_intro}

Astrometry, which deals with the accurate and precise measurement of
positions and motions of celestial objects, is the oldest branch of
observational astronomy, dating back at least to Hipparchus of Nicaea
in 190 BC. Since, from its very beginnings, this branch of astronomy
has required measurements over time to fulfill its goals, it could be
considered the precursor of the nowadays fashionable ``time-domain
astronomy''\footnote{As defined, e.g., by Wikipedia:
  https://en.wikipedia.org/wiki/Time\_domain\_astronomy}, preceding it
by at least 20 centuries. In recent years, astrometry has experienced
a ``coming of age'' motivated by the rapid increase in positional
precision allowed by the use of all-digital techniques and space
observatories (see, e.g., \citet{reff09}, Fig.~1a in \citet{Hog2017}
for an overview spanning more than 2000 years of astrometry,
\citet{Bene2017} for a summary of the contributions from the HST (fine
guide sensors), and, of course, the exquisite prospects from the
\citet{Gaia2016}, with applications ranging from fundamental
astrophysics \citep{vanaltena2013, Cacc2016}, to cosmology
\citep{lattanzi2012}).

A number of techniques have been proposed to estimate the location and
flux of celestial sources as recorded on digital detectors (CCD). In
this context, estimators based on the use of a least-squares (LS)
error principle have been widely adopted \citep{king1983accuracy,
  stetson1987daophot, alard1998method}.  The use of this type of decision rule has been
traditionally justified through heuristic reasons. First, LS methods
are conceptually straightforward to formulate based on the observation
model of these problems. Second, they offer computationally efficient
implementations and have shown reasonable performance
\citep{lee1983theoretical, stone1989, Vaki2016}. Finally, the LS
approach was the classical method used when the observations were
obtained with analog devices \citep{vanAltena_1975,euer1978}, which
are well characterized by a Gaussian noise model for the observations. In the Gaussian case the LS is equivalent to the maximum likelihood (ML) solution (\citet{Chun93},
\citet{gray_2004}, \citet{cover_2006}), and, consequently, the LS
method was taken from the analogous to the digital observational (Poisson noise model)
setting naturally.

Considering astrometry as an inference problem (of, usually, point
sources), the astrometric community has been interested for a long
time in understanding the fundamental performance limits (or
information bounds) of this task (\citet{lindegren2010} and references
therein). It is well understood by the community that the
characterization of this precision limit offers the
possibility of understanding the complexity of the task and how it
depends on key attributes of the problem, like the quality of the
observational site, the performance of the instrument (CCD), and the
details of the experimental conditions \citep{2013mendez,2014mendez}. On the other hand, it provides meaningful benchmarks to define
the optimality of practical estimators in the process of comparing
their performance with the bounds \citep{2015lobos}.

Concerning the characterization and analysis of fundamental
performance bounds, we can mention some works on the use of the
parametric \crra\ (CR) bound by
\citet{lindegren1978,jakobsen1992,zaccheo1995,adorf1996}; and
\citet{bastian2004}.  The CR bound is a minimum variance bound (MVB)
for the family of unbiased estimators
\citep{radhakrishna1945information,cramer1946contribution}.  In
astrometry and joint photometry and astrometry,
\citet{2013mendez,2014mendez} have recently studied the structure of
this bound, and have analyzed its dependency with respect to important
observational parameters under realistic (ground-based) astronomical
observing conditions. In this context, closed-form expressions for the
\crra\ bound were derived in a number of important settings (high
pixel resolution and low and high signal-to-noise ($S/N$) regimes),
and their trends were explored across different CCD pixel resolutions
and the position of the object in the CCD array. As an interesting
outcome of those studies, the analysis of the CR bound has allowed us
to predict the optimal pixel resolution of the array, as well as providing a formal justification to some heuristic
techniques commonly used to improve performance in astrometry, like
{\em dithering} for undersampled images
\citep[Sect.~3.3]{2013mendez}. Recently, an application of the CR
bound to moving sources has been done by \citet{Bouq17}, indicating
excellent agreement between our theoretical predictions, simulations,
and actual ground-based observations of the Gaia satellite, in the
context of the GBOT program (\citet{Altm14}). The use of the CR bound
on other applications is also of interest, e.g., in assessing the
performance of star trackers to guide satellites with demanding
pointing constraints \citep{Zhan2016}, or to meaningfully compare
positional differences from different catalogues (for an example
involving the SDSS and Gaia see \citet{Lemo2017}). Finally, a
formulation of the (non-parametric) Bayesian CR bound in astrometry,
using the so-called ``Van-Trees inequality'' \citep{VanT2004}, has
been presented by our group in \citet{Eche2016}: This approach is
particularly well suited for objects at the edge of detectability, and
where some prior information is available, and has been proposed for
the analysis of Gaia data for faint sources, or for those with a poor
observational history \citep{Mich2015,Mich2016}.

From the perspective of astrometric estimators, \cite{2015lobos} have
studied in detail the performance of the widely adopted LS
estimator. In particular extending the result in \citet{dsp2013},
Lobos and collaborators derived lower and upper bounds for the mean
square error (MSE) of the LS estimator. Using these bounds, the
optimality of the LS estimator was analyzed, demonstrating that for
high $S/N$ there is a considerable gap between the CR bound and the
performance of the LS estimator (indicating a lack of optimality of
this estimator). This work showed that for the
very low $S/N$ observational regime (weak astronomical sources), the
LS estimator is near optimal, as its performance closely follows the
CR bound. The limitations of the LS method in the medium to high $S/N$
regime proved in that work opens up the question of studying
alternative estimators that could achieve the CR bound on these
regimes, which is the main focus of this paper, as outlined below.

\subsection{Contribution and organization}

In this work we study the ML estimator in astrometry, motivated by its
well-known optimality properties in a classical parametric estimation setting with independent and identically distributed measurements (i.i.d.) \citep{kendall1999}. We know that in the i.i.d. case this
estimator is efficient with respect to the CR limit
\citep{kay1993fundamentals}, but it is important to emphasize that the
observational setting of astrometry deviates from the classical
i.i.d. case and, consequently, the analysis of its optimality is still
an open problem. In particular, we face the technical challenge of
evaluating its performance, a problem that, to the best of our
knowledge, has not been addressed by the astrometric
community. Concerning the independent but not identically distributed
case \citet{bradley1962} and \citet{hoadley1971} give conditions under
which ML estimators are consistent\footnote{this is, as the sample
  size increases, the sampling distribution of the estimator becomes
  increasingly concentrated at the true parameter value} and
asymptotically normal\footnote{more precisely, whose distribution
  around the true parameter approaches a normal distribution as the
  sample size grows}. Those conditions, however, are technically
difficult to proof in this context.


The main challenge here is the fact that, as in the case of the LS
estimator \citep{2015lobos}, the ML estimator is the solution of an optimization problem with a nonconvex cost function of the data. This implies
that it is not possible to directly compute the performance of the
method.  To address this technical issue, we extend the approach
proposed by \cite{fessler1996} to approximate the variance and the
mean of an implicit estimator solution of a generic optimization
problem of the data through the use of a Taylor approximation around
the mean measurement (see Theorem 1 below). Our
extension considers high order approximations of the function that
allows us not only to estimate the performance of the ML estimator
through an explicit nominal value, but also it provides a confidence
interval around it.  With this result we revisit the more general
weighted least square (WLS) and ML methods providing specific upper
and lower bounds for both methods (see Theorems~2 and
 3).  The main findings from our analysis of the bounds
are two fold: first we show that the WLS exhibits a sub-optimality similar to that of the LS method for medium to high $S/N$
regimes discovered by \cite{2015lobos} and, second, that the ML
estimator achieves the CR limit for medium to high $S/N$ and,
consequently, it is optimal on those regimes. This last result is
remarkable because, in conjunction with the result presented in
\cite{2015lobos}, we are able to identify estimators that achieve the
fundamental performance limits of astrometry in all the $S/N$ regimes
for the problem.

The paper is organized as follows: Section~\ref{sec_pre} introduces the
background, preliminaries and notation of the problem.
Section~\ref{sec_main_general} presents the main methodological
contribution of this work. Section~\ref{sec_main_astrometry} presents
the application to astrometry considering the WLS and the ML
schemes. Finally, numerical analysis of these performance bounds are
presented in Sect.~\ref{sec_empirical} and the final remarks and
conclusion are given in Sect.~\ref{final}.

\section{Preliminaries and background}
\label{sec_pre}

We begin by introducing the problem of astrometry.  For simplicity, we
focus on the 1-D scenario of a linear array detector, as it captures
the key conceptual elements of the problem\footnote{The analysis can
  be extended to the 2-D case as presented in \citet{2013mendez}.}.

\subsection{Relative astrometry as a parameter estimation problem} 
\label{sub_sec_astro_photo}

The main problem at hand is the inference of the relative position (in
the array) of a point source.  This source is modeled by two scalar
quantities, the position of object $x_c\in \mathbb{R}$ in the
array\footnote{This captures the angular position in the sky and it is
  measured in seconds of arc (arcsec thereafter), through the
  ``plate-scale'', which is an optical design feature of the
  instrument plus telescope configuration.}, and its intensity (or
brightness, or flux) that we denote by $\tilde{F}\in \mathbb{R}^+$.
These two parameters induce a probability distribution $\mu_{x_c,\tilde{F}}$ over
an observation space that we denote by $\mathbb{X}$.  Formally, given
a point source represented by the pair $(x_c,\tilde{F})$, it creates a
nominal intensity profile in a photon integrating device (PID),
typically a CCD, which can be expressed by
\begin{equation}\label{eq_pre_1}
	\tilde{F}_{x_c, \tilde{F}}(x)=\tilde{F} \cdot \phi(x-x_c,\sigma),
\end{equation}
where $\phi(x-x_c,\sigma)$ denotes the one dimensional normalized
point spread function (PSF) and where $\sigma$ is a generic parameter
that determines the width (or spread) of the light distribution on the
detector (typically a function of wavelength and the quality of the
observing site, see Sect.~\ref{sec_empirical})
\citep{2013mendez,2014mendez}.

The profile in Eq.~(\ref{eq_pre_1}) is not measured directly, but it
is observed through three sources of perturbations.  First, an
additive background noise which captures the photon emissions of the
open (diffuse) sky, and the noise of the instrument itself (the
read-out noise and dark-current \citet{Jane2001,Howe2006,Jane2007,
  McLe2008}), modeled by $\tilde{B}_i$ in
Eq.~(\ref{eq_pre_2b}). Second, an intrinsic uncertainty between the
aggregated intensity (the nominal object brightness plus the
background) and actual measurements, which is modeled by independent
random variables that follow a Poisson probability law.  Finally, we
need to account for the spatial quantization process associated with
the pixel-resolution of the PID as specified in Eqs.~(\ref{eq_pre_2b})
and (\ref{eq_pre_3}).  Modeling these effects, we have a countable
collection of independent and non-identically distributed random
variables (observations or counts) $\left\{I_i: i \in
\mathbb{Z}\right\}$, where $I_i \sim Poisson(\lambda_i(x_c,
\tilde{F}))$, driven by the expected intensity at each pixel element
$i$, given by
\begin{equation}\label{eq_pre_2b}
	\lambda_i(x_c, \tilde{F}) \equiv\mathbb{E}\{I_i\}= \underbrace{\tilde{F} \cdot
          g_i(x_c)}_{\equiv \tilde{F}_i(x_c,\tilde{F})} + \tilde{B}_i,~\forall i\in \mathbb{Z}
\end{equation}
and 
\begin{equation}\label{eq_pre_3}
	g_i(x_c) \equiv \int^{x_i+\Delta x/2}_{x_i-\Delta x/2} \phi(x- x_c,\sigma)~d x, \ \forall i \in \mathbb{Z},
\end{equation}
where $\mathbb{E}\left\lbrace \right\rbrace$ is the expectation value
of the argument and $\left\{x_i: i \in \mathbb{Z}\right\}$ denotes the standard
uniform quantization of the real line-array with resolution $\Delta
x>0$, i.e., $x_{i+1}-x_i=\Delta x$ for all $i \in \mathbb{Z}$. In
practice, the PID has a finite collection of measurement elements (or
pixels) $I_1,..,I_n$, then a basic assumption here is that we have a
good coverage of the object of interest, in the sense that for a given
position $x_c$
\begin{equation}\label{eq_pre_4}
	\sum_{i=1}^n g_i(x_c) \approx \sum_{i\in \mathbb{Z}} g_i(x_c)
        =\int_{-\infty}^{\infty} \phi(x-x_c,\sigma)~d x = 1.
\end{equation}
At the end, the likelihood (probability) of the joint observations
$I^n=(I_1,..,I_n)$ (with values in $\mathbb{N}^n$) given the source
parameters $(x_c,\tilde{F})$ is given by
\begin{equation}\label{eq_pre_5}
L({I^n; x_c,\tilde{F}}) = f_{\lambda_1(x_c,\tilde{F})}(I_1) \cdot
f_{\lambda_2(x_c,\tilde{F})}(I_2) \cdots
f_{\lambda_n(x_c,\tilde{F})}(I_n), \ \forall I^n \in \mathbb{N}^n,
\end{equation}
where $f_{\lambda}(x)=\frac{e^{-\lambda}\cdot \lambda^x}{x!}$ denotes
the probability mass function (PMF) of the Poisson law
\citep{gray_2004}.

Finally, if $\tilde{F}$ is assumed to be known\footnote{The joint
  estimation of photometry and astrometry is the task of estimating
  both $(x_c,\tilde{F})$ from the observations, see
  \cite{2014mendez}.}, the astrometric estimation is the task of
defining a decision rule $\tau_n():\mathbb{N}^n \rightarrow \Theta$,
with $\Theta=\mathbb{R}$ being the parameter space, where given an
observation $I^n$ the estimated position is given by
$\hat{x}_c(I^n)=\tau_n(I^n)$.

\subsection{The \crra\ bound} \label{subsec_cr_bounds}
In astrometry the \crra\ bound has been used to bound the variance
(estimation error) of any unbiased estimator
\citep{2013mendez,2014mendez}.  In general, let $I^n$ be a collection
of independent observations that follow a parametric PMF
$f_{\bar{\theta}}$ defined on $\mathbb{N}$. The parameters to be
estimated from $I^n$ will be denoted in general by the vector
$\bar{\theta}=(\theta_1,\theta_2,...,\theta_m) \in \Theta =
\mathbb{R}^m$.  Let $\tau_n(I^n):\mathbb{N}^n \rightarrow \Theta$ be
an unbiased estimator\footnote{In the sense that, for all
  $\bar{\theta}\in \Theta$, $\mathbb{E}_{I^n \sim f^n_{\bar{\theta}}}
  \left\lbrace \tau_n(I^n)\right\rbrace =\bar{\theta}$.}  of
$\bar{\theta}$, and $L( I^n;\bar{\theta})= f_{\bar{\theta}}(I_1)\cdot
f_{\bar{\theta}}(I_2)\cdots f_{\bar{\theta}}(I_n) $ be the likelihood
of the observation $I^n \in \mathbb{N}^n$ given $\bar{\theta}\in
\Theta$.  Then, the \crra\ bound
\citep{radhakrishna1945information,cramer1946contribution} establishes
that if
\begin{equation} \label{cond2d}
\mathbb{E}_{I^n \sim f^n_{\bar{\theta}}}\left\lbrace \frac{\partial \ln
  L(I^n; \bar{\theta}) }{\partial \theta_i} \right\rbrace = 0, \;\;  \forall i \in \left\{1,\ldots, m \right\},
\end{equation}
then, the variance (denoted by $Var$), satisfies that
\begin{equation}\label{varcr}
Var (\tau_n(I^n)_i) \geq [
  \mathcal{I}_{\bar{\theta}}(n)^{-1} ]_{i,i},
\end{equation}
where $\mathcal{I}_{\bar{\theta}}(n)$ is the {\em Fisher
information} matrix  given by
\begin{equation}\label{fisher}
[ \mathcal{I}_{\bar{\theta}}(n)]_{i,j} = \mathbb{E}_{I^n \sim f^n_{\bar{\theta}}} \left\lbrace
\frac{\partial \ln L(I^n; \bar{\theta})}{\partial \theta_i} \cdot
\frac{\partial \ln L(I^n; \bar{\theta}) }{\partial \theta_j}
\right\rbrace \;\; \forall i,j \in  \left\{1,\ldots, m \right\}.
\end{equation}
In particular, for the scalar case ($m=1$), we have that for all
$\theta \in \Theta$
\begin{equation}\label{cr_scalar}
	\min_{\tau_n(\cdot)\in \mathcal{T}^n} Var(\tau_n(I^n)) \geq \mathcal{I}_\theta(n)^{-1}=  \mathbb{E}_{I^n \sim f^n_{\theta}}  \left\lbrace\left[  \left(\frac{d \ln L(I^n; {\theta})}{d \theta} \right)^2 \right]\right\rbrace^{-1},
\end{equation}
where $\mathcal{T}^n$ is the collection of all unbiased estimators and $I^n \sim f^n_\theta$. For astrometry,
\citet{2013mendez,2014mendez} have characterized and analyzed the
\crra\ bound, leading to
\begin{proposition} (\citet[Sect.~2.4]{2014mendez}) 
If $\tilde{F} \in \mathbb{R}^+$ is fixed and known, and we want to
estimate $x_c$ from $I^n \sim f_{(x_c,\tilde{F})}=L({I^n;
  x_c,\tilde{F}})$ in
Eq.~(\ref{eq_pre_5}), then the Fisher information is given by
	\begin{equation}\label{fi_astrometry}
		\mathcal{I}_{x_c}(n) = \sum_{i=1}^n \frac{ \left(
                  \tilde{F}\frac{d g_i(x_c)}{d x_c} \right)^2
                }{\tilde{F} g_i(x_c) + \tilde{B}_i},
	\end{equation}
	which from Eq.~(\ref{cr_scalar}) induces a MVB for the
        {\em astrometric estimation problem}, and where
        $\sigma_{CR}^2(n) \equiv \mathcal{I}_{x_c}(n)^{-1}$ denotes
        the (astrometric) CR bound.
\label{CR_bound}
\end{proposition}

\subsection{Achievability and performance of the LS estimator}
\label{subsec_ls}
Concerning the achievability of the CR bound with a practical
estimator, \citet[Proposition 2]{2015lobos} have demonstrated that
this bound cannot be attained, meaning that for any unbiased estimator
$\tau_n(\cdot)$ we have that
\begin{equation}\label{eq_subsec_ls_1}
		Var(\tau_n(I^n)) > \sigma_{CR}^2, 
\end{equation}
where $I^n$ follows the Poisson PMF $f_{(x_c,\tilde{F})}$ in
Eq.~(\ref{eq_pre_5}).

This finding should be interpreted with caution, considering its pure theoretical meaning.  This is because Eq.~(\ref{eq_subsec_ls_1})
does not exclude the possibility that the CR bound could be
approximated arbitrarily close by a practical estimation scheme.
Motivated by this refined conjecture, \citet{2015lobos} proposed to
study the performance of the widely adopted LS estimator\footnote{This
  is the solution of
		$\tau_{LS}(I^n)= \arg \min_{\alpha\in \mathbb{R}}
  \sum_{i=1}^{n} \left( I_i -\lambda_i(\alpha) \right)^2$, with
  $\lambda_i(\alpha)=\tilde{F} g_i(\alpha)+\tilde{B}_i$, $\alpha$
  being a generic variable representing the astrometric position,
  $g_i(\cdot)$ is given by Eq.~(\ref{eq_pre_3}), and where $\arg \min$ represents the argument that minimizes the expression. More details are
  presented in \citet{2015lobos}.}  with the goal of deriving
operational upper and lower performance bounds of its performance that
could be used to determine how far could this scheme depart from the
CR limit. Then, from this result, it was possible to evaluate the
goodness of the LS estimator for concrete observational regimes.  For
bounding the performance of the LS estimator, the challenge was that
$\tau_{LS}(I^n)$ is an implicit function of the data (where no
close-form expression is available) and, consequently,
\citet{2015lobos} derived a result to bound the estimation error and
the variance of $\tau_{LS}(I^n)$.  We can briefly summarize the main
result presented in \citet[Theorem 1]{2015lobos} saying that under
certain mild sufficient conditions (that were shown to be realistic
for astrometry), there is a constant $\delta>0$ (that depends on the
observational regime, in particular the $S/N$) and a nominal variance
$\sigma^2_{LS}$, which is determined in closed-form in the result,
from which it is possible to bound $Var(\tau_{LS}(I^n))$ by the simple
expression
\begin{equation}\label{eq_subsec_ls_2}
Var(\tau_{LS}(I^n))  \in \left( \frac{\sigma^2_{LS}(n)}{(1+\delta)^2}, 
                \frac{\sigma^2_{LS}(n)}{(1-\delta)^2} \right), 
\end{equation}
where
\begin{equation}\label{eq_subsec_ls_3}
\sigma^2_{LS}(n) = 
\frac{\sum_{i=1}^n (\tilde{F}g_i(x_c)+\tilde{B}_i) \cdot
  (g_i'(x_c))^2}{\left(\tilde{F}\sum_{i=1}^n (g_i'(x_c))^2\right)^2}.
\end{equation}
Note that when $\delta$ is small, $\sigma^2_{LS}(n)$ tightly
determines the performance of the LS estimator, and its comparison
with $\sigma^2_{CR}$ can be used to evaluate the goodness of the LS
estimator for astrometry. Based on a careful comparison, it was shown
in \citet[Sect.~4]{2015lobos} that, in general, $\sigma^2_{LS}(n)$ is
close to $\sigma^2_{CR}(n)$ for the small $S/N$ regime of the problem.
However for moderate to high $S/N$ regimes, the gap between
$\sigma^2_{LS}(n)$ and $\sigma^2_{CR}(n)$ becomes quite
significant\footnote{In particular, for the very high $S/N$ regime and
  assuming $\Delta x/\sigma \ll 1$, \citet[Proposition 3]{2015lobos}
  shows that this gap reaches the condition
$\frac{\sigma^2_{LS}(n)}{\sigma_{CR}^2(n)} \approx \frac{8}{3\sqrt{3}}$.}.

These unfavorable findings for the LS method have motivated us to
study alternatives schemes that could potentially approach better the \crra\ bound for the rich observational context of
medium to high $S/N$ regimes.  This will be the focus of the following
sections, where in particular we explore the performance of the ML and
WLS estimators, thus extending and generalizing the analysis done for
the LS estimator by our group presented in \cite{2015lobos}.

\section{Bounding the performance of an implicit estimator}
\label{sec_main_general}
Before we go to the case of the WLS and the ML estimators, we present
a general result that bounds the performance of {\it any} estimator that is
the solution of a generic optimization problem. Let us consider a
vector of observations $I^n=(I_1,\ldots, I_n) \in \mathbb{R}^n$ and a
general so-called cost function $J(\alpha,I^n)$. Then the estimation
of $x_c$ from the data is the solution of the following optimization
problem:
\begin{equation} \label{eq_main_general_1}
	\tau_J(I^n) \equiv \arg \min_{\alpha \in \mathbb{R}} J(\alpha,I^n),
\end{equation}
where $\alpha$ represents the position of the object in the context of
astrometry.  As in our previous work \citep{2015lobos}, the challenge
here is that this estimator is implicit because no closed-form
expression of the data which solves Eq.~(\ref{eq_main_general_1}) is
assumed.  In particular, this implies that both the variance and the
estimation error of $\tau_J(I^n)$ can not be determined directly.  To
address this technical issue, we extend the approach proposed by
\citet{fessler1996} to approximate the variance and the mean of an
implicit estimator solution of a problem described by
Eq.~(\ref{eq_main_general_1}) through the use of a Taylor
approximation around the mean measurement, i.e.,
$\bar{I}^n=\mathbb{E}_{I^n\sim f_{(x_c,\tilde{F})}}(I^n)$.

More precisely, we assume that $J(\alpha,I^n)$ has a unique global
minimum at $\tau_J(I^n)$, and that it has a regular behavior, so its
partial derivatives are zero, i.e.,
 \begin{equation}\label{eq_main_general_2}
 0=\left.\frac{\partial}{\partial
   \alpha}J(\alpha,I^n)\right|_{\alpha=\tau_J(I^n)}\equiv
 \frac{\partial}{\partial \alpha}J(\tau_J(I^n),I^n).
 \end{equation}
Then we can obtain $\tau_J(I^n)$ by a first order Taylor expansion
around the mean $\bar{I}^n$ by
\begin{equation}\label{eq_main_general_3}
\tau_J(I^n)=\tau_J(\bar{I}^n)+\sum_{i=1}^n\frac{\partial}{\partial
  I_i}\tau_J(\bar{I}^n)(I_i-\bar{I}_i)+\underbrace{\frac{1}{2}\sum_{i=1}^n\sum_{j=1}^n\frac{\partial^2}{\partial
    I_i\partial
    I_j}\tau_J(\bar{I}^n-t(I^n-\bar{I}^n))(I_i-\bar{I}_i)(I_j-\bar{I}_j)}_{\equiv
  e(\bar{I},I-\bar{I})}.
\end{equation} 
with $t \in [0,1]$ is fixed but unknown\footnote{It follows that
  $\lim_{I\rightarrow \bar{I}}\limits \frac{e(\bar{I},I-\bar{I})}{||
    I-\bar{I}||_2}=0$.}.  For simplicity,
Eq.~(\ref{eq_main_general_3}) can be written in matrix form as
\begin{equation}\label{eq_main_general_4}
\tau_J(I^n)= \tau_J(\bar{I}^n)+\nabla\tau_J(\bar{I}^n)\cdot (I^n-\bar{I}^n)+ e(\bar{I}^n,I^n-\bar{I}^n),
\end{equation}
where $\nabla=[\frac{\partial}{\partial
    I_1}\ldots\frac{\partial}{\partial I_n}]$ denotes the row gradient
operator and $e(\bar{I}^n,I-\bar{I}^n)$ is the residual error of the
Taylor expansion. From Eq.~(\ref{eq_main_general_4}) we can
  readily obtain the following expression for its variance
\begin{align}\label{eq_main_general_5}
Var\{\tau_J(I^n)\} &= \underbrace{\nabla\tau_J(\bar{I}^n)Cov\{I ^n \}\nabla\tau_J(\bar{I}^n)^T}_{\equiv \sigma^2_{J}(n) }+\underbrace{Var\{ e(\bar{I}^n,I^n-\bar{I}^n) \}+2Cov\{\nabla\tau_J(\bar{I}^n) (I^n-\bar{I}^n),e(\bar{I}^n,I^n-\bar{I}^n)\}}_{\equiv \gamma_{J} (n)}.
\end{align}
In Eq.~(\ref{eq_main_general_5}) we recognize two terms:
$\sigma^2_{J}(n)$ that captures the linear behaviour of
$\tau_J(\cdot)$ around $\bar{I}^n$ and $\gamma_{J} (n)$ which reflects
the deviation from this linear trend.  It should be noted that the
above expression does not depend on $\tau(I^n)$ itself, but on its
partial derivatives evaluated at the mean vector of observations. Then
in the adoption of this approach to estimate $Var\{\tau_J(I^n)\} $, a
key task is to determine $\nabla\tau_J(\bar{I}^n)$.
\begin{remark}
It is meaningful to note that \citet{fessler1996} only considered the
linear term in his approximate analysis, obviating the residual term
$\gamma_{J} (n)$ in Eq.~(\ref{eq_main_general_5}). This first order
reduction is not realistic for our problem because the solution of a
problem like the one posed by Eq.~(\ref{eq_main_general_1}) in
astrometry has important non-linear components that need to be
considered in the analysis of Eq.~(\ref{eq_main_general_5}).
\end{remark}

In an effort to analyze both the linear and non-linear aspects of a
general intrinsic estimator solution to Eq.~(\ref{eq_main_general_1}),
the following result offers sufficient conditions to determine
$\sigma^2_{J}(n)$ in closed-form, and to bound the magnitude of the
residual term $\gamma_{J} (n)$ in Eq.~(\ref{eq_main_general_5}).

\begin{theorem}
\label{th_main_general}
Let us consider a fixed and unknown parameter $x_c \in \R$, the
observations $I^n=(I_1,...,I_n)^T$ where $I_i \sim f_{x_c}$, and
$\tau_J(I^n)$ the estimator solution of
Eq.~(\ref{eq_main_general_1}). If we satisfy the following two 
  rather general conditions:
\begin{itemize}
\item[a)]the cost function $J(\alpha, I^n)$ is twice differentiable
  with respect to $I^n$ and $x_c$, and the gradient of $\tau_J(\cdot)$
  evaluated in the mean data $\bar{I}^n$
offers the following decomposition
\begin{equation}
\nabla \tau_J(\bar{I}^n) \cdot (I^n-\bar{I}^n)=a\sum_{i=1}^N\limits b_i(I_i-\bar{I}_i)
\label{close_gradient}
\end{equation}
with $a$ and $\{b_i: i \in \{1,..., N\}\}$ constants, and,
\item[b)] the estimator evaluated in the mean data equals the true
  parameter $x_c \in \R$, this is,
\begin{equation}
\tau_J(\bar{I}^n) =x_c, 
\label{noise_free}
\end{equation}
\end{itemize}
then we can define two new quantities $\epsilon_{J} (n)$ and
$\beta_{J} (n)$ (both $> 0$) and $\sigma^2_{J} (n)$ in
Eq.~(\ref{eq_main_general_5}) with analytical expressions (details
presented in Appendix \ref{proof_th_main_general}) such that
\begin{equation}
| \underbrace{\mathbb{E}_{I^n \sim f_{x_c}}\{ \tau_J(I^n)\}-x_c }_{\text{bias}} | \leq \epsilon_{J} (n)
\label{bias_eqn}
\end{equation}
and
\begin{equation}
Var_{I^n \sim f_{x_c}}\{\tau_J(I^n)\} \in  \left ( \sigma^2_{J}(n) - \beta_{J} (n) , \sigma^2_{J}(n) + \beta_{J} (n)  \right).
\label{VAR_eqn}
\end{equation}
\end{theorem}
The proof of this result and the expression for $(\epsilon_{J} (n),
\sigma^2_{J} (n), \beta_{J} (n))$ in Eqs.~(\ref{bias_eqn}) and
(\ref{VAR_eqn}) are presented in detail in Appendix
\ref{proof_th_main_general}.

Revisiting the equality in Eq.~(\ref{eq_main_general_5}), Theorem
  1 provides general sufficient conditions to
bound the residual term $\gamma_{J} (n)$ and by doing that, a way of
bounding
the variance of $\tau_J(I^n)$ which is the solution of
Eq.~(\ref{eq_main_general_1}). In particular, it is worth noting that
if the ratio $\frac{\beta_{J} (n)}{\sigma^2_{J}(n)} \ll 1$, then
Eq.~(\ref{VAR_eqn}) offers a tight bound for $Var_{I^n \sim
  f_{x_c}}\{\tau_J(I^n)\}$.
In this last context, $\sigma^2_{J}(n)$ (called the nominal value of
the result) provides a very good approximation for $Var_{I^n \sim
  f_{x_c}}\{\tau_J(I^n)\}$.

On the application of this result to the WLS and ML estimators, we
will see that the main assumption in Eq.~(\ref{close_gradient}) is
satisfied in both cases (see Eqs.~(\ref{eq_proof_th_wls_5})
and~(\ref{cov_bound_2}) in Appendix~\ref{proof_th_wls} and
\ref{proof_th_ml}, respectively), and from that
$\sigma^2_{J}(n)$ is playing an important role to approximate the
performance of ML and WLS in a wide range of observational regimes. In
addition, the analysis of the bias in Eq.~(\ref{bias_eqn}) shows that
these estimators are unbiased for any practical purpose and,
consequently, contrasting their performance (estimation error $\sim$
$\sqrt{\text{variance}}$) with the CR bound is a meaningful way to
evaluate optimality.

\section{Application to astrometry}
\label{sec_main_astrometry}
In this section we apply Theorem 1 to bound
the variances of the ML and WLS estimators in the context of
astrometry.  Following the model presented in
Sect.~\ref{sub_sec_astro_photo}, $I^n=(I_1,\ldots,I_n)^T$ denotes the
measurements acquired by each pixel of the array, and where each of
them follows a Poisson distribution given by
\begin{equation}\label{eq_sec_main_astrometry_1}
I_i\sim Poisson (\lambda_i(x_c)),~~ i=1,\ldots, n, 
\end{equation}  
as expressed by Eqs.~(\ref{eq_pre_2b}) and (\ref{eq_pre_3}).

\subsection{Bounding the variance of the WLS estimator}
\label{sub_sec_wls}
The WLS estimator, denoted by $\tau_{WLS}(I^n)$ in
Eq.~(\ref{eq_sec_main_astrometry_3}), is implicitly defined through a
cost function given by
\begin{equation}
J_{WLS}(\alpha,I^n)=\sum_{i=1}^nw_i(I_i-\lambda_i(\alpha))^2,
\label{WLS_cost_fn}
\end{equation}
where $(w_1,\ldots,w_n)^T~\in\mathbb{R}^n_+$ is a weight vector, and
$\alpha$ is a general source position parameter. Specifically we have
that
\begin{equation}\label{eq_sec_main_astrometry_3}
\tau_{WLS}(I^n) = \arg \min_{\alpha \in\mathbb{R}}J_{WLS}(\alpha,I^n).
\end{equation}
Applying Theorem 1 we obtain the following
result:

\begin{theorem}\label{th_wls}
Let us consider the WLS estimator solution of
Eq.~(\ref{eq_sec_main_astrometry_3}), then we have that

\begin{equation}\label{eq_sec_main_astrometry_4}
| \underbrace{\mathbb{E}_{I^n \sim f_{x_c}}\{ \tau_{WLS}(I^n)\}-x_c }_{\text{bias}} | \leq \epsilon_{WLS} (n)
\end{equation}
and 
\begin{equation}\label{eq_sec_main_astrometry_5}
Var_{I^n \sim f_{x_c}}\{\tau_{WLS}(I^n)\} \in  \left ( \sigma^2_{WLS}(n) - \beta_{WLS} (n) , \sigma^2_{WLS}(n) + \beta_{WLS} (n)  \right),
\end{equation}
where $\sigma^2_{WLS}(n)$ is given by
\begin{eqnarray} \label{eq_sec_main_astrometry_6}
\sigma^2_{WLS}(n)  &=& \frac{\sum_{i=1}^nw_i^2\lambda_i(x_c)\left.\left(\frac{\partial \lambda_i(\alpha)}{\partial \alpha}\right)^2\right|_{\alpha =x_c}}{\left(\sum_{i=1}^nw_i\left.\left(\frac{\partial \lambda_i(\alpha)}{\partial \alpha}\right)^2\right|_{\alpha =x_c}\right)^2}
\end{eqnarray}
and $\beta_{WLS}(n)$ and $\epsilon_{WLS}(n)$ are well defined analytical expression of the problem (presented in Appendix \ref{proof_th_wls}).  
\end{theorem}

The proof of this result and the expressions for $\epsilon_{WLS}(n)$
and $\beta_{WLS}(n)$in Eqs.~(\ref{eq_sec_main_astrometry_4}) and
(\ref{eq_sec_main_astrometry_5}), respectively, are elaborated in
Appendix \ref{proof_th_wls}. This result offers concrete expressions to bound the bias as well as
the variance of the WLS estimator. For the bias bound in
Eq.~(\ref{eq_sec_main_astrometry_4}), it will be shown that
$\epsilon_{WLS} (n)$ is very small (order of magnitudes smaller than
$x_c$) for all the observational regimes explored in this work and,
consequently, the WLS can be considered an unbiased estimator in
astrometry, as it would be expected. Concerning the bounds for the
variance in Eq.~(\ref{eq_sec_main_astrometry_5}), we will show that
for high and moderate $S/N$ regimes the ratio $\beta_{WLS} (n)
/\sigma^2_{WLS}(n)\ll 1$ and consequently in this context
$\sigma^2_{WLS}(n)$ is a precise estimator of $Var_{I^n \sim
  f_{x_c}}\{\tau_{WLS}(I^n)\}$. For the very small $S/N$ the results
offers an admissible interval $\sigma^2_{WLS}(n) \pm \beta_{WLS} (n)$
around the nominal value $\sigma^2_{WLS}(n)$. Therefore in any context
$\sigma^2_{WLS}(n)$ shows to be a meaningful approximation for the
performance of the WLS.
	
	\begin{remark}
	\label{key_remark_wls}
	If we focus on the analysis of the closed form expression
        $\sigma^2_{WLS}(n)$ as an approximation of $Var_{I^n \sim
          f_{x_c}}\{\tau_{WLS}(I^n)\}$ and we compare it with the CR
        bound $\sigma^2_{CR}(n)$ in Eqs.~(\ref{fi_astrometry}) and
        (\ref{eq_subsec_ls_1}), we note that they are very similar in
        their structure.  In particular, it follows that
        $\sigma^2_{WLS}(n)=\sigma^2_{CR}(n)$, if an only if, the
        weights of the WLS estimator are selected in the following way
	\begin{equation}\label{eq_sec_main_astrometry_7}
		w_i=K \cdot \frac{1}{\lambda_i(x_c)},~~\forall i\in\{1,\ldots,n\},
	\end{equation}
where $K$ is an arbitrary constant ($K > 0$). In other words, the only
way in which the performance of the WLS approximates the CR limit is
if we select the weights as in Eq.~(\ref{eq_sec_main_astrometry_7}).
However, this selection uses the information of the true position
$x_c$, which is unfeasible as it contradicts the very essence of the
inference task (indeed, $x_c$ is unknown, and we are trying to
  estimate it from the data).  Another interpretation is that no
matter how we choose the weights of the WLS estimator, it is not
possible that the WLS is close to the CR bound for every position
$x_c$, telling us that the WLS is intrinsically not optimal from the
perspective of being close to the CR limit in all the possible
astrometric scenarios. In particular, this impossibility result is
very strong in the high $S/N$ regimes where $\sigma^2_{WLS}(n)\approx
Var_{I^n \sim f_{x_c}}\{\tau_{WLS}(I^n)\}$.  This implication is
consistent with the analysis presented by \citet[Fig.~4]{2015lobos},
where it was shown that the variance of the LS estimator is
significantly higher than then CR bound in the high $S/N$ regime. This justifies the study of the ML estimator.
	\end{remark}

\subsection{Bounding the variance of the ML estimator}
\label{sub_sec_ml}
The ML estimator, denoted by $\tau_{ML}(I^n)$ in
Eq.~(\ref{ML_cost_fn}), is implicitly defined through a cost function
\begin{equation}\label{eq_sec_main_astrometry_8}
J(\alpha,I^n)=\sum_{i=1}^nI_i
\ln(\lambda_i(\alpha))-\lambda_i(\alpha),
\end{equation}
where $\alpha$ is a general source position parameter. Specifically,
given an observation $I^n$ we have that
\begin{eqnarray}
\tau_{ML}(I^n)&=&\arg
\max_{\alpha\in\mathbb{R}}J(\alpha,I^n),\nonumber\\
&=&\arg \min_{\alpha\in\mathbb{R}}\sum_{i=1}^n-I_i
\ln(\lambda_i(\alpha))+\lambda_i(\alpha).
\label{ML_cost_fn}
\end{eqnarray}
Applying Theorem 1 we obtain the following
result:

\begin{theorem}\label{th_ml}
Let us consider the ML estimator solution of Eq.~(\ref{ML_cost_fn}),
then we have that
\begin{equation}\label{eq_sec_main_astrometry_10}
| \underbrace{\mathbb{E}_{I^n \sim f_{x_c}}\{ \tau_{ML}(I^n)\}-x_c }_{\text{bias}} | \leq \epsilon_{ML} (n)
\end{equation}
and 
\begin{equation}\label{eq_sec_main_astrometry_11}
Var_{I^n \sim f_{x_c}}\{\tau_{ML}(I^n)\} \in  \left ( \sigma^2_{ML}(n) - \beta_{ML} (n) , \sigma^2_{ML}(n) + \beta_{ML} (n)  \right),
\end{equation}
where
\begin{eqnarray} \label{eq_sec_main_astrometry_12}
\sigma^2_{ML}(n) = \sigma^2_{CR}(n)= \left( \sum_{i=1}^n \frac{ \left(
  \tilde{F}\frac{d g_i(x_c)}{d x_c} \right)^2 }{\tilde{F} g_i(x_c) +
  \tilde{B}_i} \right)^{-1},
\end{eqnarray}
and $\beta_{ML}(n)$ and $\epsilon_{ML}(n)$ are well defined analytical
expression of the problem (presented in Appendix \ref{proof_th_ml}).
\end{theorem}
The proof of this result and the expressions for $\epsilon_{ML}(n)$
and $\beta_{ML}(n)$ in Eqs.~(\ref{eq_sec_main_astrometry_10}) and
(\ref{eq_sec_main_astrometry_11}), respectively, are elaborated in
Appendix \ref{proof_th_ml}.

\begin{remark}
	It is important to mention that the magnitude of
        $\epsilon_{ML} (n)$ is orders of magnitude smaller than $x_c$
        in all the observational regimes studied in this work (see
        this analysis in Sect.~\ref{sec_empirical}) and, consequently,
        for any practical purpose the ML is an unbiased
        estimator. This implies that the comparison with the CR bound
        is a meaningful indicator when evaluating the optimality of the
        ML estimator.
\end{remark}

\begin{remark}
We observe that if the ratio $\beta_{ML} (n) /\sigma^2_{ML}(n)$ is
significantly smaller than one, which is shown in
Sect.~\ref{sec_empirical} from medium to high $S/N$ regimes, then
$Var_{I^n \sim f_{x_c}}\{\tau_{ML}(I^n)\} \approx
\sigma^2_{ML}(n)$. This is a very interesting result because we can
approximate the performance of the ML estimator with
$\sigma^2_{ML}(n)$. On this context, it is remarkable to have that the
nominal value $\sigma^2_{ML}(n)$ is precisely the CR bound (see
Eq.~(\ref{eq_sec_main_astrometry_12})), because this means that the ML
estimator closely approximate this MVB in the interesting regime from
moderate to very high $S/N$.  Note that this medium-high $S/N$ regime
is precisely the context where the LS estimator shows significant
deficiencies as presented in \citet{2015lobos}. Therefore, ML offers
optimal performances in the regime where LS type of methods are not
able to match the CR bound, which satisfactorily resolves the question
posted by \citet{2015lobos} on the study of schemes that could closely approach the CR bound in the high $S/N$ regime.
\end{remark}

\section{Numerical analysis}
\label{sec_empirical}
In this section we evaluate numerically the performance bounds
obtained in Sect.~\ref{sec_main_astrometry} for the WLS and ML
estimators, and compare them with the astrometric CR bound in Proposition 1. The idea is to consider some realistic
astrometric conditions to evaluate the expressions developed in 
  Theorems~2 and 3 and their dependency on
important observational conditions and regimes. As we shall see,
key variables in this analysis are the tradeoff between the intensity
of the object and the noise represented by the $S/N$ value, and the
pixel resolution of the CCD.

\subsection{Experimental setting}
\label{subsec_setting}
We adopt some realistic design and observing variables to model the
problem \citep{2013mendez,2014mendez}. For the PSF, analytical and
semi-empirical forms have been introduced, see for instance the
ground-based model in \cite{king1971} and the space-based models by
\cite{king1983accuracy} or \cite{bendinelli1987}. In this work we will
adopt a Gaussian PSF, i.e., $\phi(x,\sigma)= \frac{1}{\sqrt{2\pi}
  \sigma} e^{- \frac{(x)^2}{2 \sigma^2}}$ in Eq.~(\ref{eq_pre_3}), and
where $\sigma$ is the width of the PSF and is assumed to be
known. This PSF has been found to be a good representation for typical
astrometric-quality Ground-based data \citep{2010mendez}. In terms of
nomenclature, $FWHM \equiv 2 \sqrt{2 \ln 2} \,\, \sigma$ measured in
arcsec, denotes the {\em Full-Width at Half-Maximum} parameter, which
is an overall indicator of the image quality at the observing site
\citep{chromey2010measure}.
  
The background profile, represented by $ \left\{\tilde{B}_i,
i=1,..,n\right\}$ in Eq.~(\ref{eq_pre_2b}), is a function of several
variables, like the wavelength of the observations, the moon phase
(which contributes significantly to the diffuse sky background), the
quality of the observing site, and the specifications of the
instrument itself.  We will consider a uniform background across
pixels underneath the PSF, i.e., $\tilde{B}_i=\tilde{B}$ for all
$i$. To characterize the magnitude of $\tilde{B}$, it is important to
first mention that the detector does not measure photon counts $[e^-]$
directly, but a discrete variable in ``{\em Analog to Digital Units}
(ADUs)'' of the instrument, which is a linear proportion of the photon
counts (\cite{McLe2008}). This linear proportion is characterized by
the gain of the instrument $G$ in units of $[e^-/ADU]$.  $G$ is just a
scaling value, where we can define $F \equiv \tilde{F}/G$ and $B
\equiv \tilde{B}/G$ as the intensity of the object and noise,
respectively, in the specific ADUs of the instrument.  Then, the
background (in {ADUs}) depends on the pixel size $\Delta x$~arcsec as
follows
\begin{equation}\label{eq_sec_empirical_1}
B=f_s\Delta x+ \frac{D+RON^2}{G}~ [ADU],
\end{equation}
where $f_s$ is the (diffuse) sky background in ADU~arcsec$^{-1}$, while
$D$ and $RON^2$, both measured in e$^{-}$ model the dark-current and
read-out-noise of the detector on each pixel, respectively. Note that
the first component in Eq.~(\ref{eq_sec_empirical_1}) is attributed to
the site, and its effect is proportional to the pixel size.  On the
other hand, the second component is attributed to errors of the PID
(detector), and it is pixel-size independent. This distinction is
central when analyzing the performance as a function of the pixel
resolution of the array (see details in
\citet[Sect.~4]{2013mendez}). More important is the fact that in
typical ground-based astronomical observation, long exposure times are
considered, which implies that the background is dominated by diffuse
light coming from the sky, and not from the detector
\cite[Sect.~4]{2013mendez}.

For the experimental conditions, we consider the scenario of a
ground-based station located at a good site with clear atmospheric
conditions and the specification of current science-grade CCDs, where
$f_s=1502.5$~ADU~arcsec$^{-1}$, $D=0$, $RON=5$~e$^-$, $FWHM=1$~arcsec
(equivalent to $\sigma= 1/(2\sqrt{2\ln 2})$~arcsec) and
$G=2$~e$^-$~ADU$^{-1}$ (with these values we will have $B=313$~ADU for
$\Delta x=0.2$~arcsec using Eq.~(\ref{eq_sec_empirical_1})). In terms
of scenarios of analysis, we explore different pixel resolutions for
the CCD array $\Delta x\in [0.1,0.65]$ measured in arcsec, and
different signal strengths $\tilde{F} \in \left\{1080,
3224,20004,60160\right\}$, measured in e$^-$, which corresponds to
$S/N$ $\in \sim \{12,32,120,230 \}$.  Note that increasing $\tilde{F}$
implies increasing the $S/N$ of the problem, which can be
approximately measured by the ratio $\tilde{F}/\tilde{B}$. On a given
detector plus telescope setting, these different $S/N$ scenarios can
be obtained by appropriately changing the exposure time (open shutter)
that generates the image.

\subsection{Bias analysis}
\begin{figure}[ht] 
  \begin{minipage}[b]{0.5\linewidth}
  \centering
    \includegraphics[width=1.05\linewidth]{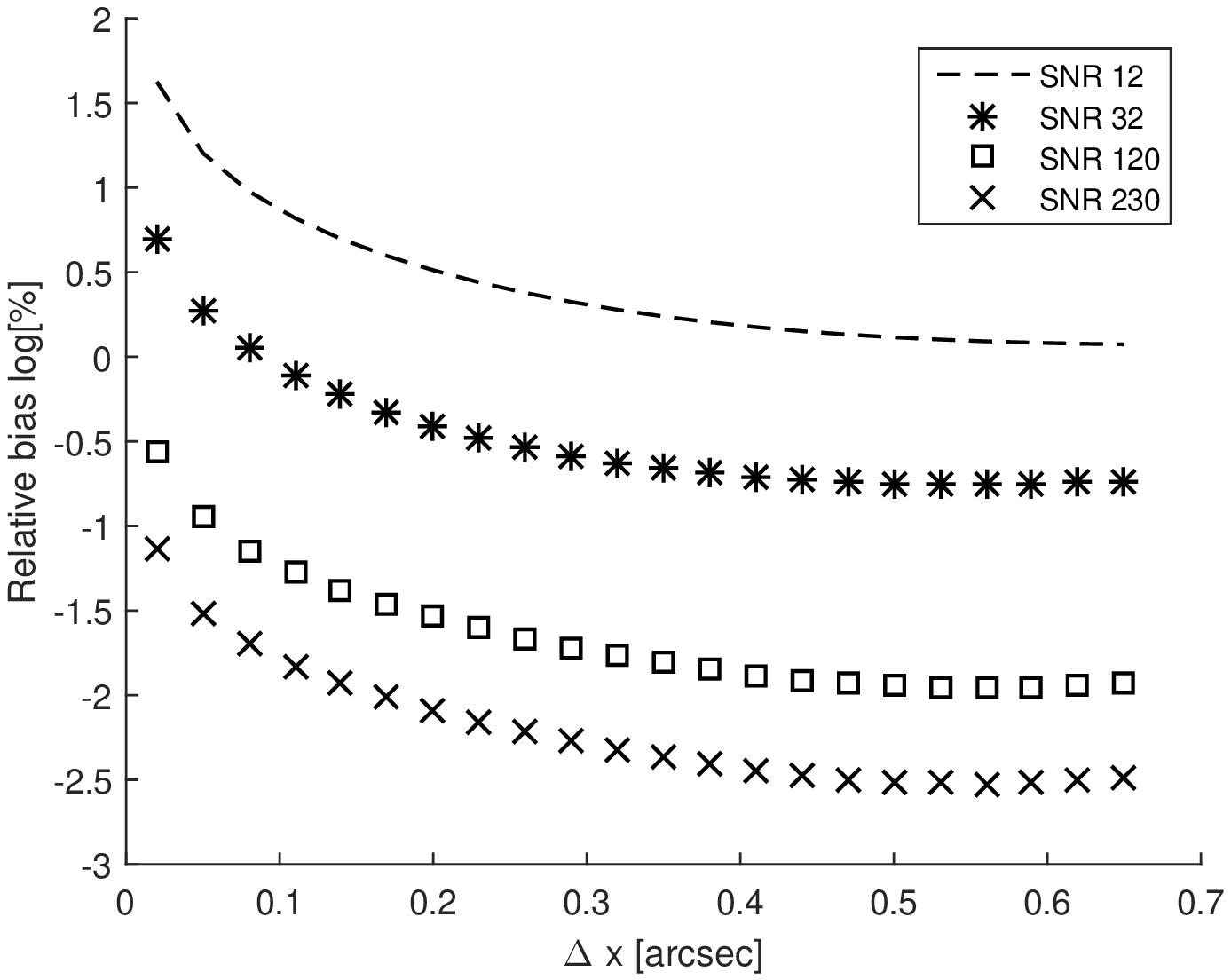} 
    \label{fig6:a} 
  \end{minipage} 
  \begin{minipage}[b]{0.5\linewidth}
    \centering
    \includegraphics[width=1.05\linewidth]{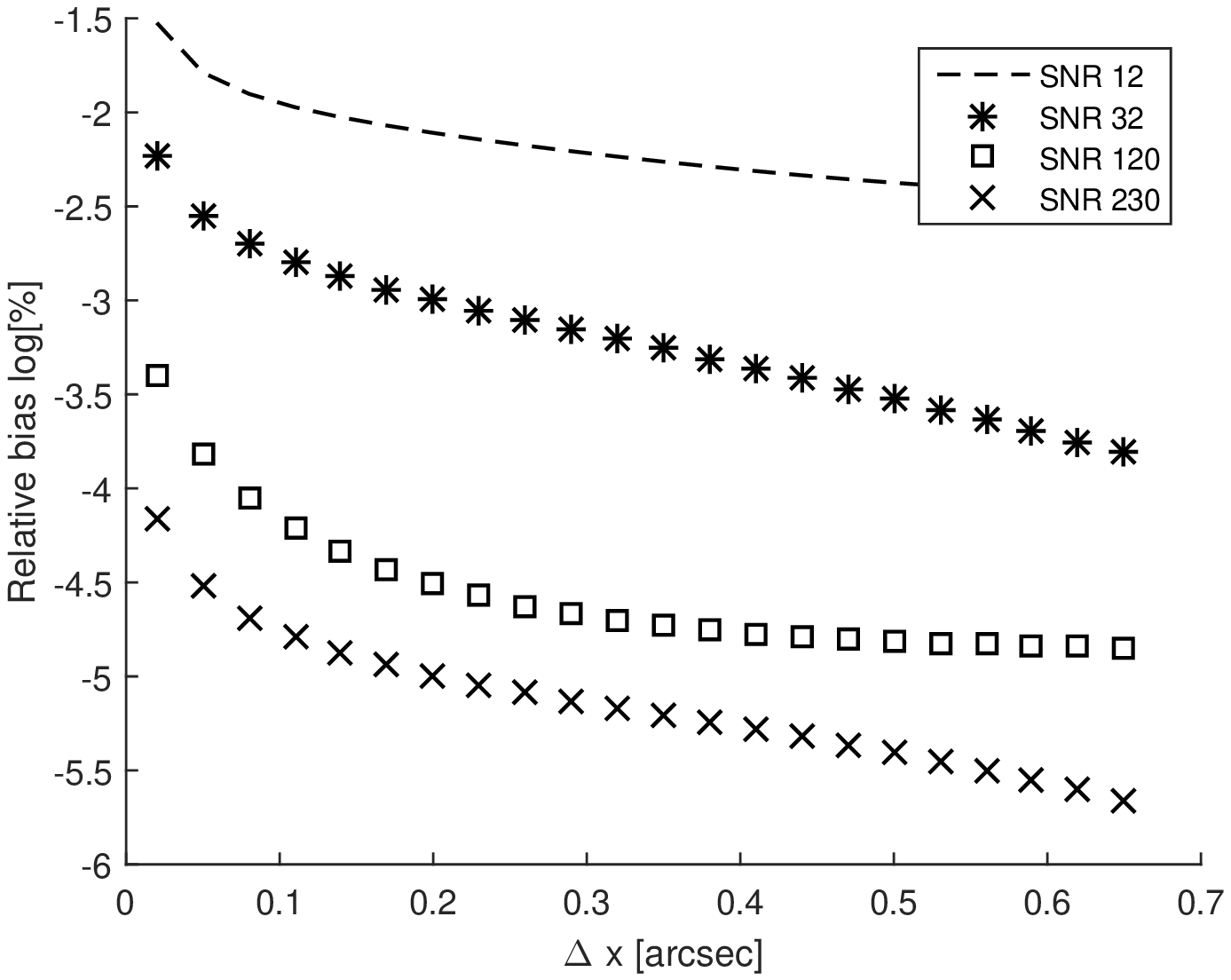} 
      \label{fig6:b} 
  \end{minipage} 
  \caption{Relative performance of the bias (as measured by
      $\log\left( 100 \times \frac{\epsilon_J(n)}{x_c} \right)$)
    stipulated by  Theorem 1 for the WLS estimator (left side,
    Eq.~(\ref{eq_sec_main_astrometry_4})) and the ML estimator (right
    side, Eq.~(\ref{eq_sec_main_astrometry_10})). Results are reported
    for different values of the source flux $\tilde{F} \in
    \left\{1080, 3224,20004,60160\right\}$, all in e$^-$ (top to
    bottom symbols respectively), as a function of the detector pixel
    size. The $0\%$ level corresponds to having achieved no bias.}
   \label{WLS_bias_relative}
\end{figure}

Considering the upper bound terms $\epsilon_{WLS} (n)$ and
$\epsilon_{ML} (n)$ for the bias error obtained from Theorems
  2 and 3 for the WLS and ML, respectively,
Fig.~\ref{WLS_bias_relative} presents the relative bias error for
different $S/N$ regimes and pixel resolutions.  In the case of the ML
estimator, the bounds for relative bias error are very small in all
the explored $S/N$ regimes and pixel resolutions meaning that for any
practical purposes this estimator is unbiased as expected from theory
\citep{gray_2004}. For the case of the WLS, we observe that from
medium to high $S/N$ the relative error bound obtained is very small
but, meanwhile at low $S/N$, unbiasedness can not be fully guaranteed
from the bound in Eq.~(\ref{eq_sec_main_astrometry_4}).  In general,
our results show that both WLS and ML are unbiased estimators for
astrometry in a wide range of relevant observational regimes (in
particular from medium to high $S/N$) and, consequently, it is
meaningful to analyze the optimality of these estimators in comparison
with the CR bound in those regimes.

In the following sections, we move to the analysis of the variance of
the WLS and ML with particular focus on the medium to high $S/N$
regimes across all pixel resolutions using the performance bounds
derived in Eqs.~(\ref{eq_sec_main_astrometry_5})
and~(\ref{eq_sec_main_astrometry_11}), respectively.

\subsection{Performance analysis of the WLS estimator}
\label{subsec_anal_wls}
In this section, we evaluate numerically the expression derived in
Theorem 2 to bound the variance of the WLS estimator
in Eq.~(\ref{eq_sec_main_astrometry_5}). For that we characterize the
admissible regime predicted for the variance of the WLS estimator,
i.e., the interval
\begin{equation}\label{eq_subsec_anal_wls_1}
\left ( \sigma^2_{WLS}(n) - \beta_{WLS} (n) , \sigma^2_{WLS}(n) + \beta_{WLS} (n)  \right), 
\end{equation}
for $S/N$ $\in \{12,32,120,230 \}$ and $\Delta x \in
[0.01,0.65]$~arcsec.  In these bounds, we recognize its central value
(or nominal value) $\sigma^2_{WLS}(n)$ in
Eq.~(\ref{eq_sec_main_astrometry_6}) and the length of the interval
$2\beta_{WLS} (n)$ that is determined in closed form for its numerical
evaluation in Eqs.~(\ref{eq_proof_th_main_general_2}) and
(\ref{eq_proof_th_wls_8}). Note that $2\beta_{WLS} (n)$ can be
considered an indicator of the precision of our result to approximate
the variance of the WLS in astrometry.
\begin{figure}
  \begin{minipage}[b]{0.5\linewidth}
  \centering
    \includegraphics[width=1.05\linewidth]{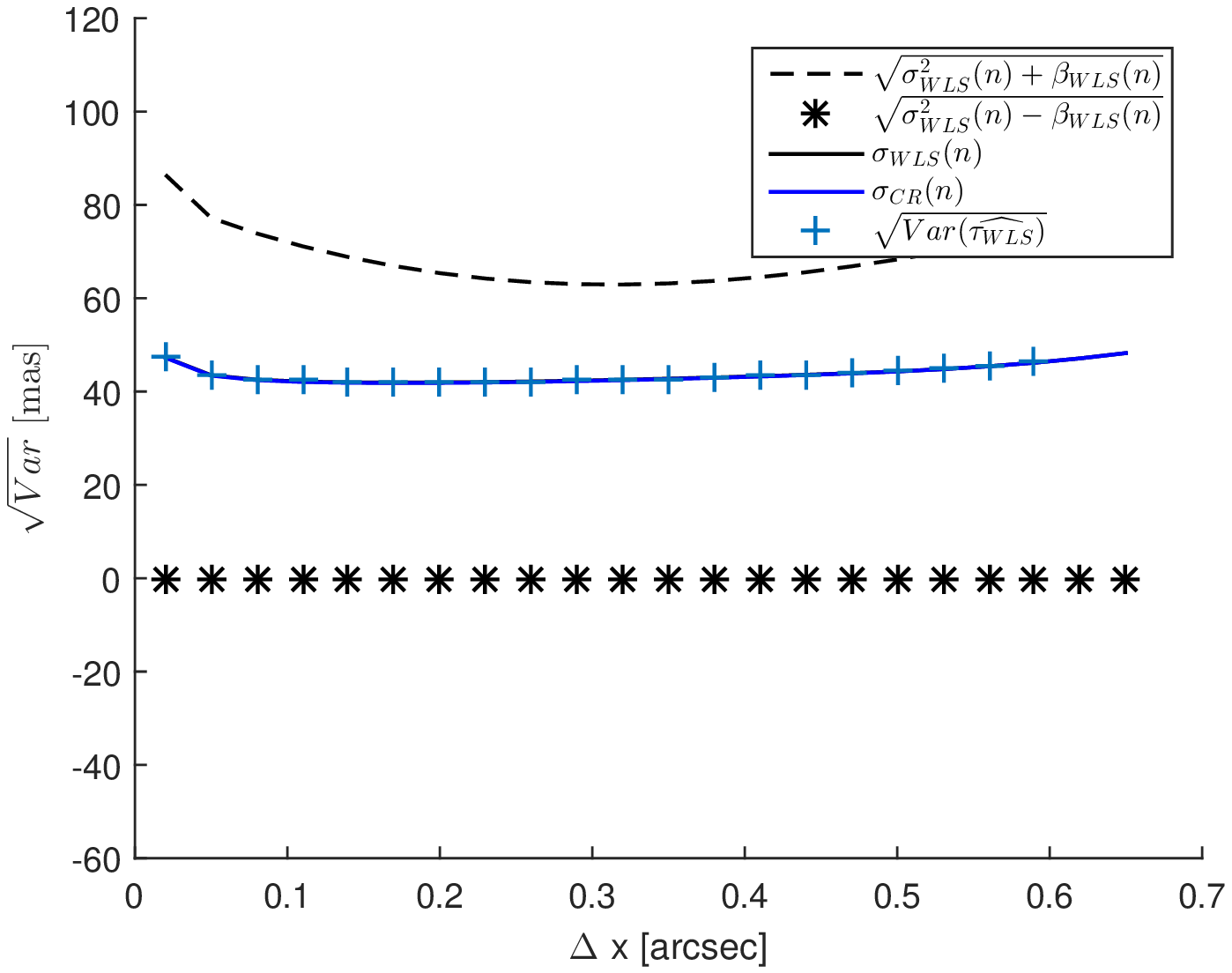} 
    \label{fig4:a} 
  \end{minipage} 
  \begin{minipage}[b]{0.5\linewidth}
    \centering
    \includegraphics[width=1.05\linewidth]{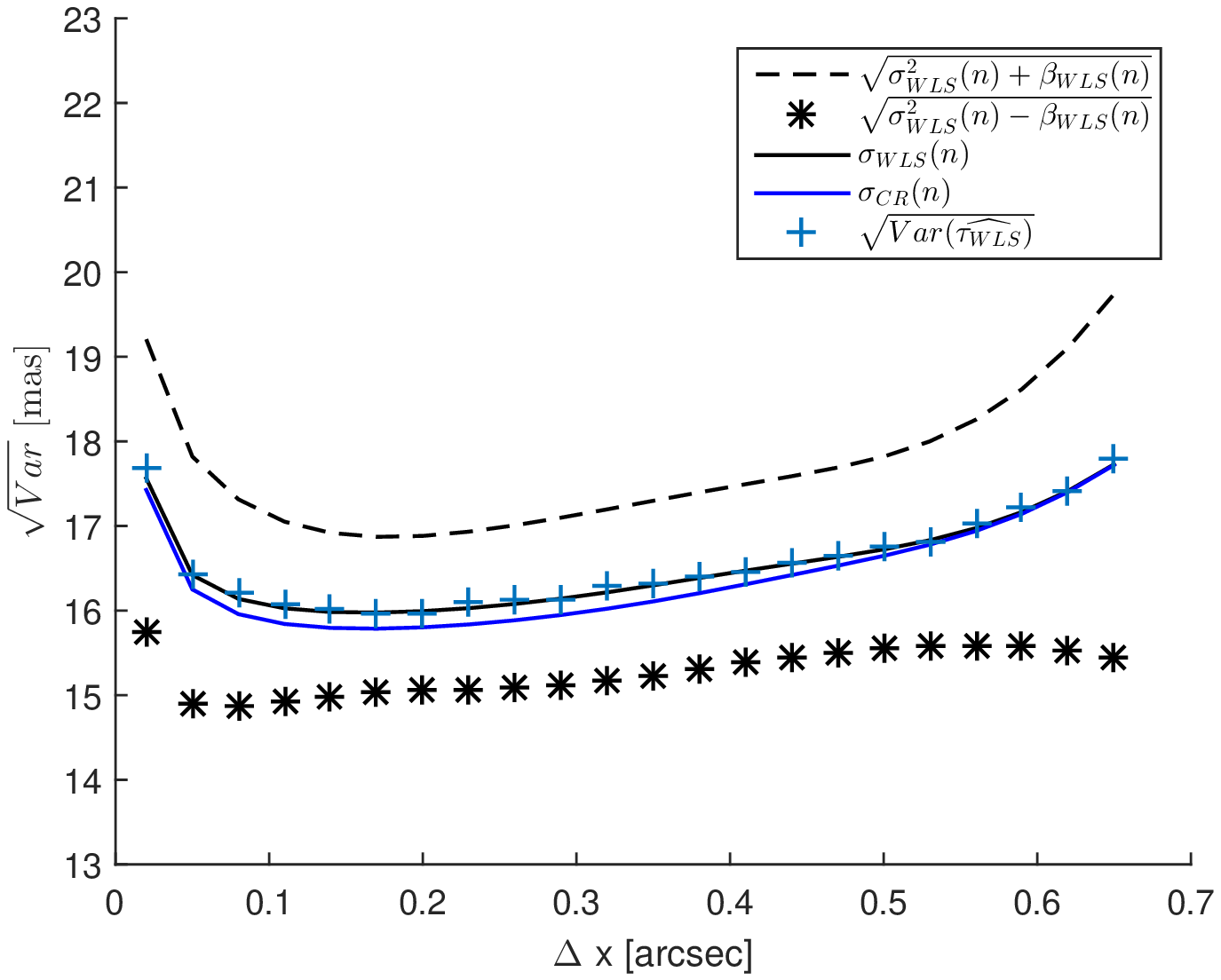} 
      \label{fig4:b} 
  \end{minipage} 
  \begin{minipage}[b]{0.5\linewidth}
    \centering
    \includegraphics[width=1.05\linewidth]{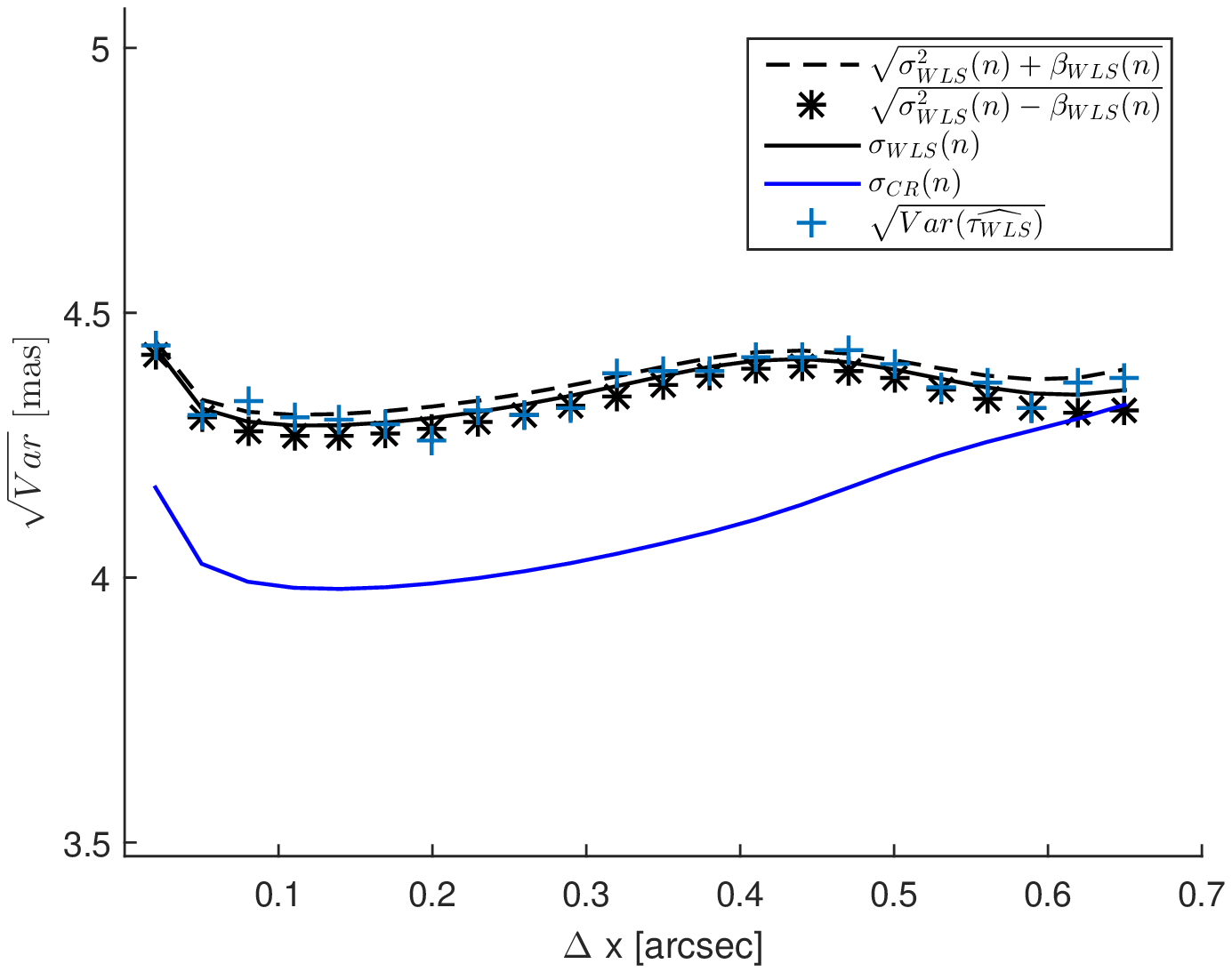} 
      \label{fig4:c}  
  \end{minipage}
  \hfill
  \begin{minipage}[b]{0.5\linewidth}
    \centering
    \includegraphics[width=1.05\linewidth]{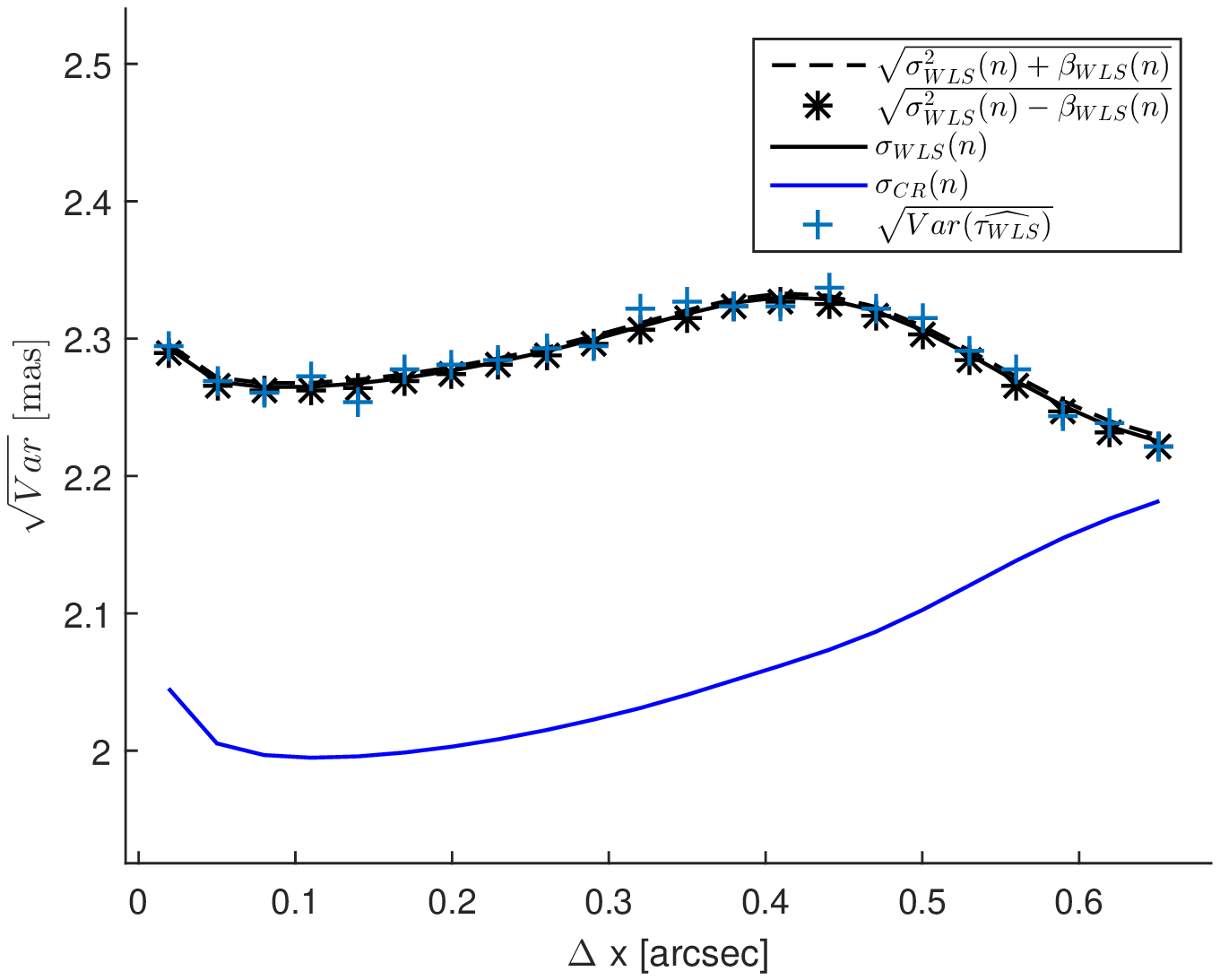} 
      \label{fig4:d} 
  \end{minipage} 
  \caption{Range of the square root of the variance performance (in
    miliarcsecond=mas) for the WLS method in astrometry using uniform weights (equivalent to the LS method) predicted by
    Theorems 1 and 2,
      Eq.~(\ref{eq_sec_main_astrometry_5}). Results are reported for
    different representative values of $\tilde{F}$ and across
    different pixel sizes (top-left to bottom-right): $\tilde{F} \in
    \left\{1080, 3224,20004,60160\right\}$~e$^-$.}
   \label{WLS_v}
\end{figure}

\subsubsection{Revisiting the uniform weight case}
\label{uniform_weight_sect}
To begin the analysis, we consider the setting of uniform weights
across pixels, i.e., the case of the LS estimator and, without loss of
generality, we locate the object in the center of the field of
view\footnote{It is important to remind that the CR bound is a
  function of the value of the parameter to be estimated, in this case
  the position $x_c$, see the Fisher information in
  Eq.~(\ref{fi_astrometry}).}, which can be considered a reasonable
scenario to represent the complexity of the astrometry task.
At this point, it is important to remind the reader that from the
analysis of $\sigma^2_{WLS}(n)$ in Remark 2,
the nominal value $\sigma^2_{WLS}(n)$ is equivalent to the CR bound
when the $w_i$ are selected as a function of the true position in
Eq.~(\ref{eq_sec_main_astrometry_7}). In view of this observation, the
selection of non-uniform weights can be interpreted as biasing the
estimation to a particular area of the angular space, which goes in
contradiction with the essence of the inference problem that estimates
the position with no prior information, and only relies on the
measured counts. From this interpretation, revisiting the LS estimator
is an important first step in the analysis of the WLS framework.

On the specifics, the boundaries of the interval in
Eq.~(\ref{eq_subsec_anal_wls_1}) and its nominal values are
illustrated in Fig.~\ref{WLS_v} for the different observational
regimes. In addition, Fig.~\ref{WLS_v} shows the CR bound as a
reference to evaluate the optimality of the LS scheme across settings.
We observe that for the low $S/N \sim 12$ regime, the nominal values
precisely match the CR bound, however our result is not conclusive as
the interval around the nominal performance is significantly large.
This is the regime where our result is not conclusive regarding the
performance of the LS estimator. In the regime $S/N \in (30, 50)$ (top
right panel on Fig.~\ref{WLS_v}), we notice an important reduction on
the range of admissible performance, and our result becomes more
informative and meaningful. In this context, the nominal values is
very close to the CR bound, and we could assert that the LS estimator
offers sufficiently good performance in the sense that is very
competitive with the MVB.  Importantly, when we move to
the regime of relatively high $S/N$ and very hight $S/N$ (from $100$
to $300$), our results are very accurate to predict the performance of
the LS method, and we find that the gap between the CR and the nominal
value is very significant (the deviation from the MVB is
$16\%$ and $30\%$ for $S/N$ $120$ and $230$ at $\Delta x=0.2$~arcsec,
respectively). This last result confirms one of the main findings
presented in \citet{2015lobos}, who showed that for medium to high
$S/N$ the LS estimator is suboptimal with respect to the MVB.  In Fig.~\ref{WLS_v} we also show the square root of
  the empirical variance ($Var(\hat{\tau_{WLS}})$) with respect to the
  empirically-determined mean position $\hat x_c$ (using the WLS
  estimator), all as derived from the simulations, showing good
  consistency with our predictions.

\subsubsection{Non-uniform weight case}
The sub-optimality of the WLS scheme from moderate to high $S/N$ seen in Fig. \ref{WLS_v} can
be extended for any arbitrary selection of a fixed set of weights, as
it would be expected. Given that the space of weights selection is
literally unlimited, we use the insight obtained from Remark
  2 that states that a selection of weights can be
interpreted as an specific prior on the position of the object where
the optimum, but unfortunately unknown, selection (achieving the CR
bound) of weights in Eq.~(\ref{eq_sec_main_astrometry_7}) is an
explicit function of the unknown position of the object.  Then, we
consider a finite set of positions $\left\{x_{c,1},..,x_{c,M}
\right\}$ that uniformly partition the field of view, and their
corresponding weights sets $\left\{w_i(x_{c,1}): 1=1,..,n \right\}$,
$\left\{w_i(x_{c,2}): 1=1,..,n \right\} \ldots$ and
$\left\{w_i(x_{c,M}): 1=1,..,n \right\}$ using
Eq.~(\ref{eq_sec_main_astrometry_7}) to cover a representative
collections of weights for the problem of astrometry.

Then, for a specific selection of weights in our admissible collection
(attributed to a prior belief of the position of the object in the
field of view), we evaluate the worse discrepancy between
  $\sigma^2_{WLS}(n) - \beta_{WLS} (n)$ (which is the most favourable
  expression for the variance predicted from 
    Theorem~2), and the CR bound in 
    Proposition~1, across a collection of presumed object
  positions in the following range of positions
$$\Theta= \{x_o^{\ast}-\sigma , x_o^{\ast}-0.8*\sigma
  ,x_o^{\ast}-0.6*\sigma ,x_o^{\ast}-0.4*\sigma,
  x_o^{\ast}-0.2*\sigma, x_o^{\ast} \}$$
where $x_o^{\ast}$ denotes the center of the array (which, as indicated at the beginning of Sect. \ref{uniform_weight_sect}, is equal to the true object position $x_c$)  and
$\sigma=FWHM/2\sqrt{2\ln 2}$ is the dispersion parameter of the PSF.
The idea of using this worse case difference is justified from the
fact that in this parameter estimation problem we do not know the
position of the object, and consequently, the optimality of any WLS
estimator should be evaluated in the worse case situation, as the scheme
should be able to estimate the position of the object in any scenario
(position). More precisely, for a given weight selection
$\left\{w_i,i=1,.., n \right\}$, we use the following worse case
discrepancy
\begin{equation}\label{eq_subsec_anal_wls_2}
\sup_{x_c \in \Theta } \frac{(\sigma^2_{WLS}(n) - \beta_{WLS} (n)) -
  \sigma^2_{CR}(n)}{\sigma^2_{CR}(n)}.
\end{equation}
For this analysis note that both $(\sigma^2_{WLS}(n) - \beta_{WLS}
(n))$ and $\sigma^2_{CR}(n)$ are functions of the position $x_c$,
$\Delta x$, and $S/N$.

\begin{figure} [t]
  \begin{minipage}[b]{0.5\linewidth}
  \centering
    \includegraphics[width=1.05\linewidth]{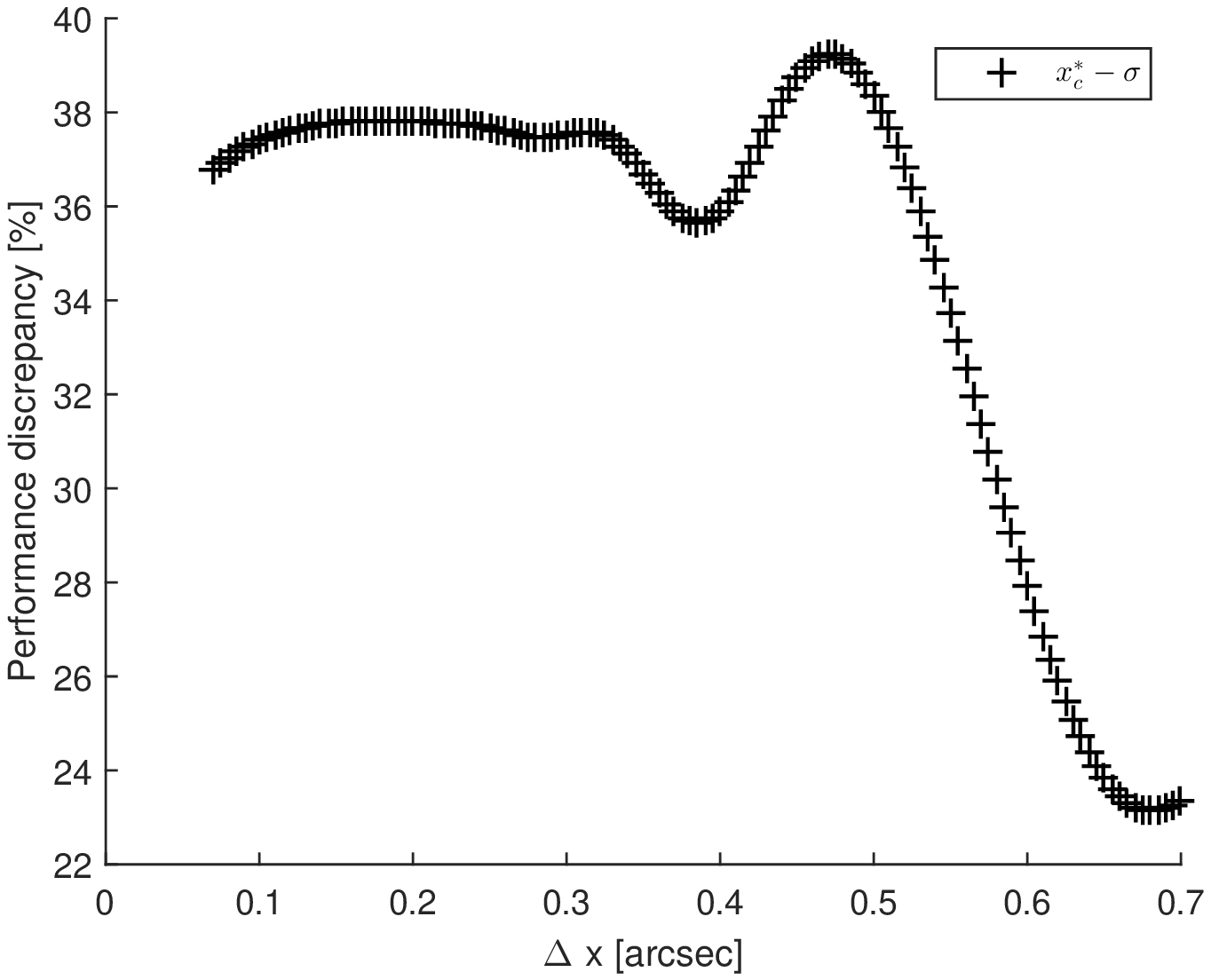} 
  \end{minipage} 
  \begin{minipage}[b]{0.5\linewidth}
    \centering
    \includegraphics[width=1.05\linewidth]{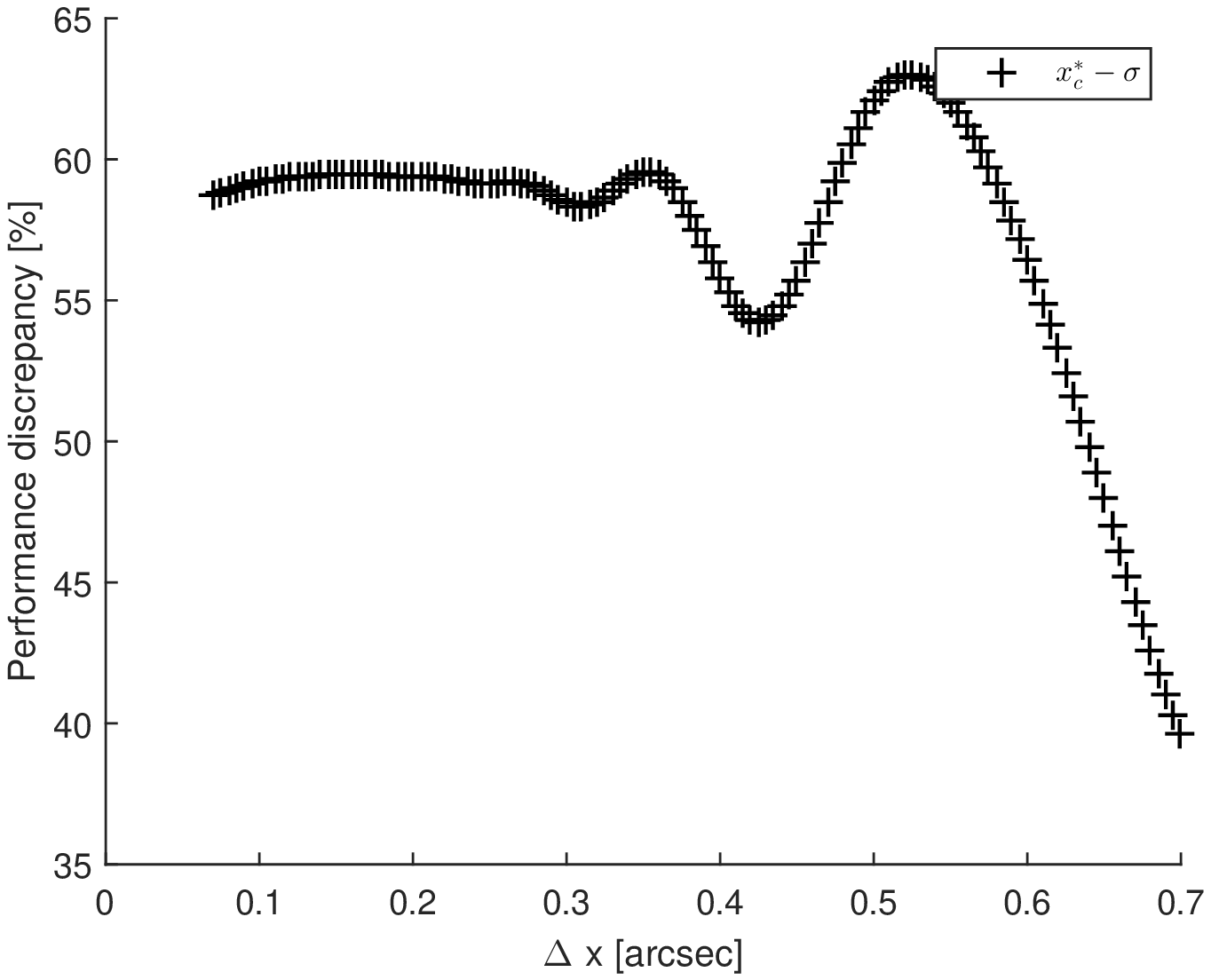} 
  \end{minipage} 
 \caption{Worse case discrepancies in Eq.~(\ref{eq_subsec_anal_wls_2})
   for the WLS estimator using the weights set indexed by the
   positions $\Theta= \{x_o^{\ast}-\sigma , x_o^{\ast}-0.8*\sigma
   ,x_o^{\ast}-0.6*\sigma ,x_o^{\ast}-0.4*\sigma,
   x_o^{\ast}-0.2*\sigma, x_o^{\ast} \}$. Results are reported for two
   $S/N$ scenarios, namely $\tilde{F}=20004$~e$^-$ (Left) and
   $\tilde{F}=60160$~e$^-$ (Right), and across different pixel sizes.}
   \label{WLS_weightworsecase}
\end{figure}

Fig.~\ref{WLS_weightworsecase} illustrates the worse case discrepancy
in Eq.~(\ref{eq_subsec_anal_wls_2})
for the medium and high $S/N$ regimes where Theorem~2
provides an accurate and meaningful prediction of the performance of
the WLS method, i.e., for $S/N$ $\in \{120,230 \}$, and across $\Delta
x \in [0.05,0.7]$~arcsec.  The discrepancy is quite significant, on the order of
$37\%$ and $60\%$ in the range for $\Delta x \in [0.1,0.3]$~arcsec for
$S/N$~$120$ and $230$, respectively.
\begin{figure} [t]
  \begin{minipage}[b]{0.5\linewidth}
  \centering
    \includegraphics[width=1.05\linewidth]{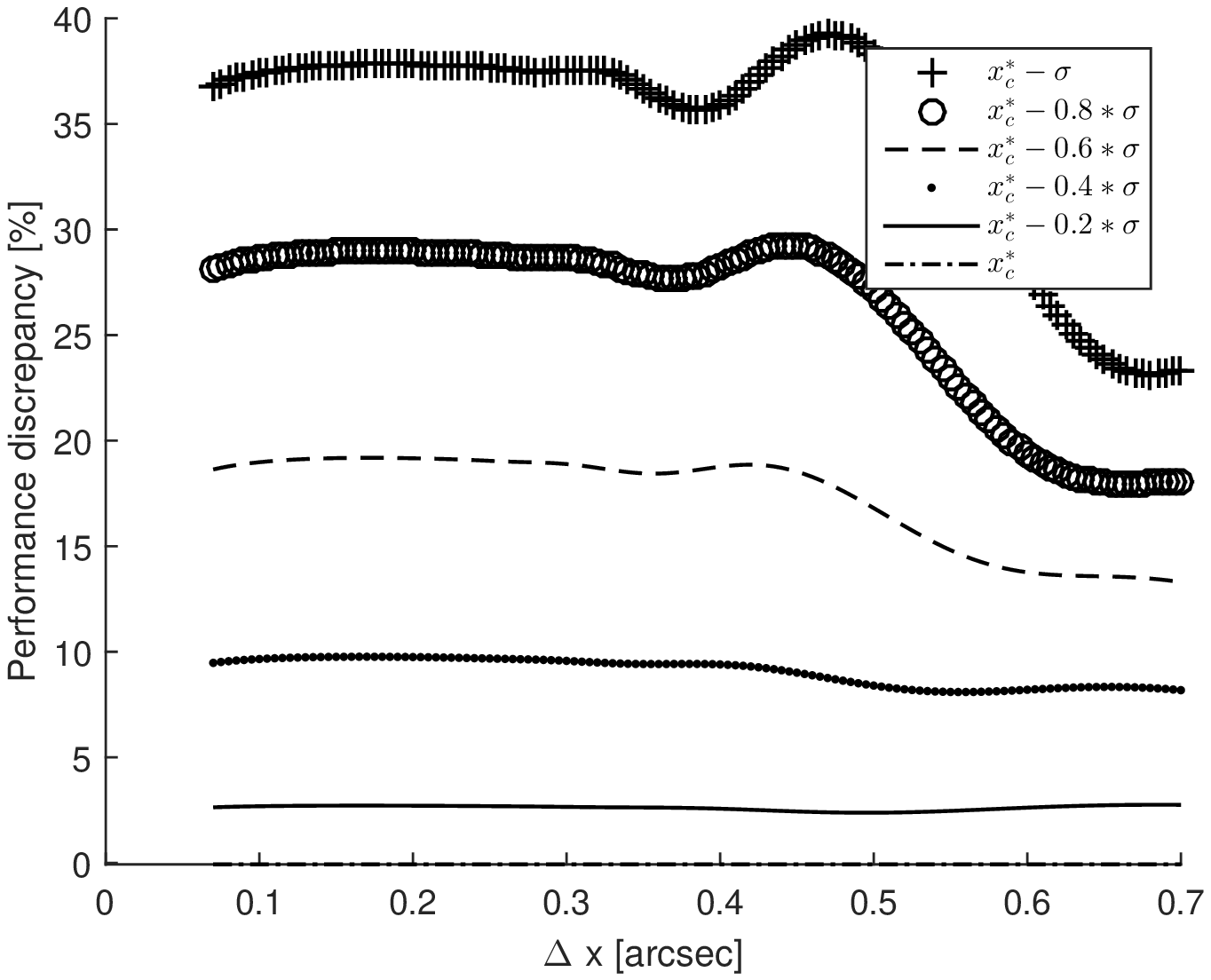} 
  \end{minipage} 
  \begin{minipage}[b]{0.5\linewidth}
    \centering
    \includegraphics[width=1.05\linewidth]{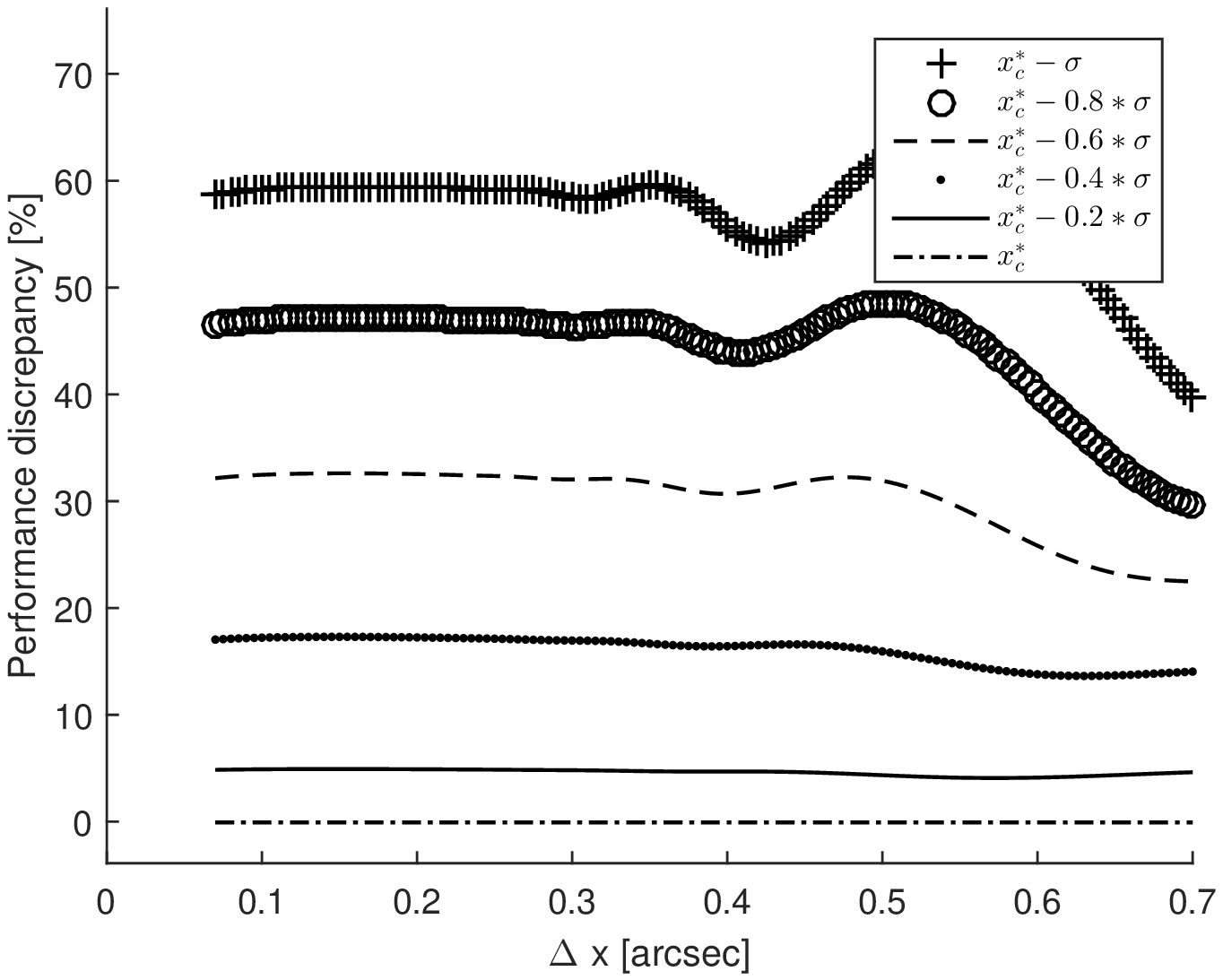} 
  \end{minipage} 
 \caption{Performance discrepancies (measuring the non-optimality) of
   the WLS estimator using the center position as a prior for the
   weight selection with respect to the CR bound obtained for the
   true object positions $\{x_o^{\ast}-\sigma , x_o^{\ast}-0.8*\sigma
   ,x_o^{\ast}-0.6*\sigma ,x_o^{\ast}-0.4*\sigma,
   x_o^{\ast}-0.2*\sigma, x_o^{\ast} \}$. Results are reported for two
   $S/N$ scenarios, namely $\tilde{F}=20004$~e$^-$ (Left) and
   $\tilde{F}=60160$~e$^-$ (Right), and across different pixel sizes.}
   \label{WLS_weight}
\end{figure}

To refine the worse case analysis presented in
Fig.~\ref{WLS_weightworsecase}, and to evaluate in more detail the
sensitivity of the discrepancy indicator given by
Eq.~(\ref{eq_subsec_anal_wls_2}), we evaluate the discrepancy of WLS
using the weights associated to $x_o^{\ast}$ (the center position of
the array) with respect to the CR bound associated to the positions
$\{x_o^{\ast}-\sigma , x_o^{\ast}-0.8*\sigma ,x_o^{\ast}-0.6*\sigma
,x_o^{\ast}-0.4*\sigma, x_o^{\ast}-0.2*\sigma, x_o^{\ast} \}$, to
study how the discrepancy (measuring the non-optimality of the method)
evolves when the adopted position moves far from the prior imposed by
the WLS in the center of the array. Fig.~\ref{WLS_weight} illustrates
this behaviour, where it is possible to see that the discrepancy is
very sensitive and proportional to the misassumption of the object
position, where the worse case discrepancy in the maximum
  achievable location precision is on the order of $40\%$ for pixel
sizes in the range $[0.1,0.6]$~arcsec for $S/N \sim 120$, and about
$60\%$ for pixel sizes in the range $[0.1,0.6]$~arcsec for $S/N \sim
230$.  These worse case scenario happens in both cases when the object
is located the farthest from the prior assumption, as it would be
expected.

The main conclusion derived form this CR bound analysis is that,
independent of the weight selection adopted, as long as the weights
are fixed, the WLS estimator is not able to achieve the CR bound in
all observational regimes. More precisely, the discrepancy (measuring
the non-optimality) in the less favourable case of an hypothetical and
feasible position of the object is very significant, in the range of
$40\%-60\%$ for the important regime of high and very high $S/N$.

\subsection{Performance analysis of the ML estimator}
\label{subsec_anal_ml}
In this section we perform the same analysis done for the WLS in
Sect.~\ref{subsec_anal_wls}, but using the result in 
  Theorem~3.  In particular, we consider the admissible
regime for the variance of the ML estimator given by
$$\left ( \sigma^2_{ML}(n) - \beta_{ML} (n) , \sigma^2_{ML}(n) +
\beta_{ML} (n) \right)$$ in Eq.~(\ref{eq_sec_main_astrometry_11}),
where the nominal value in this case, $\sigma^2_{ML}(n)$ in
Eq.~(\ref{eq_sec_main_astrometry_12}), is precisely the CR bound,
while the length of the interval $2\beta_{ML} (n)$ is given by
Eqs.~(\ref{eq_proof_th_main_general_2}) and (\ref{ml_identity}) in
closed form.

\begin{figure}[tbp] 
  \begin{minipage}[b]{0.5\linewidth}
  \centering
    \includegraphics[width=1.05\linewidth]{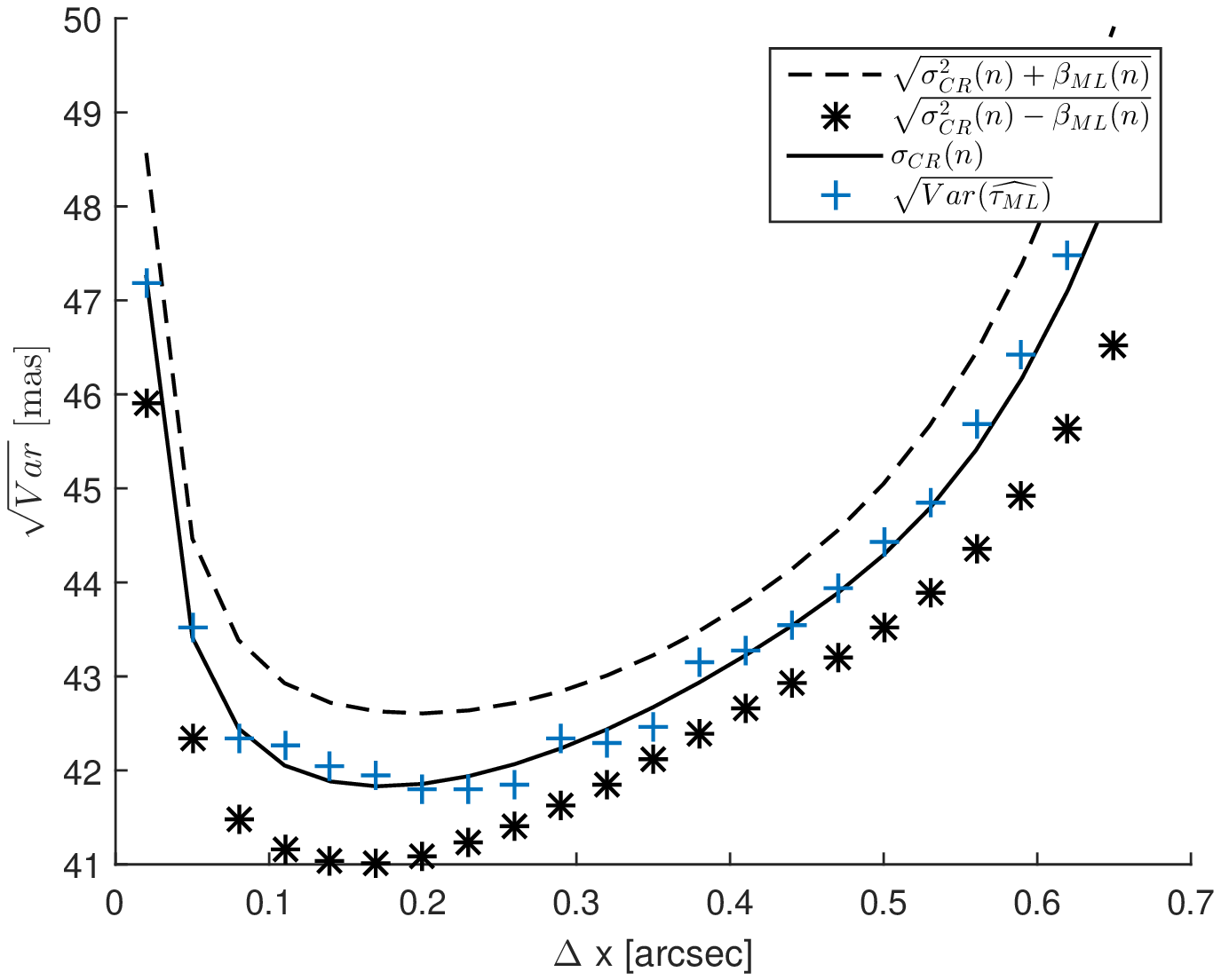} 
    \label{fig2:a} 
  \end{minipage} 
  \begin{minipage}[b]{0.5\linewidth}
    \centering
    \includegraphics[width=1.05\linewidth]{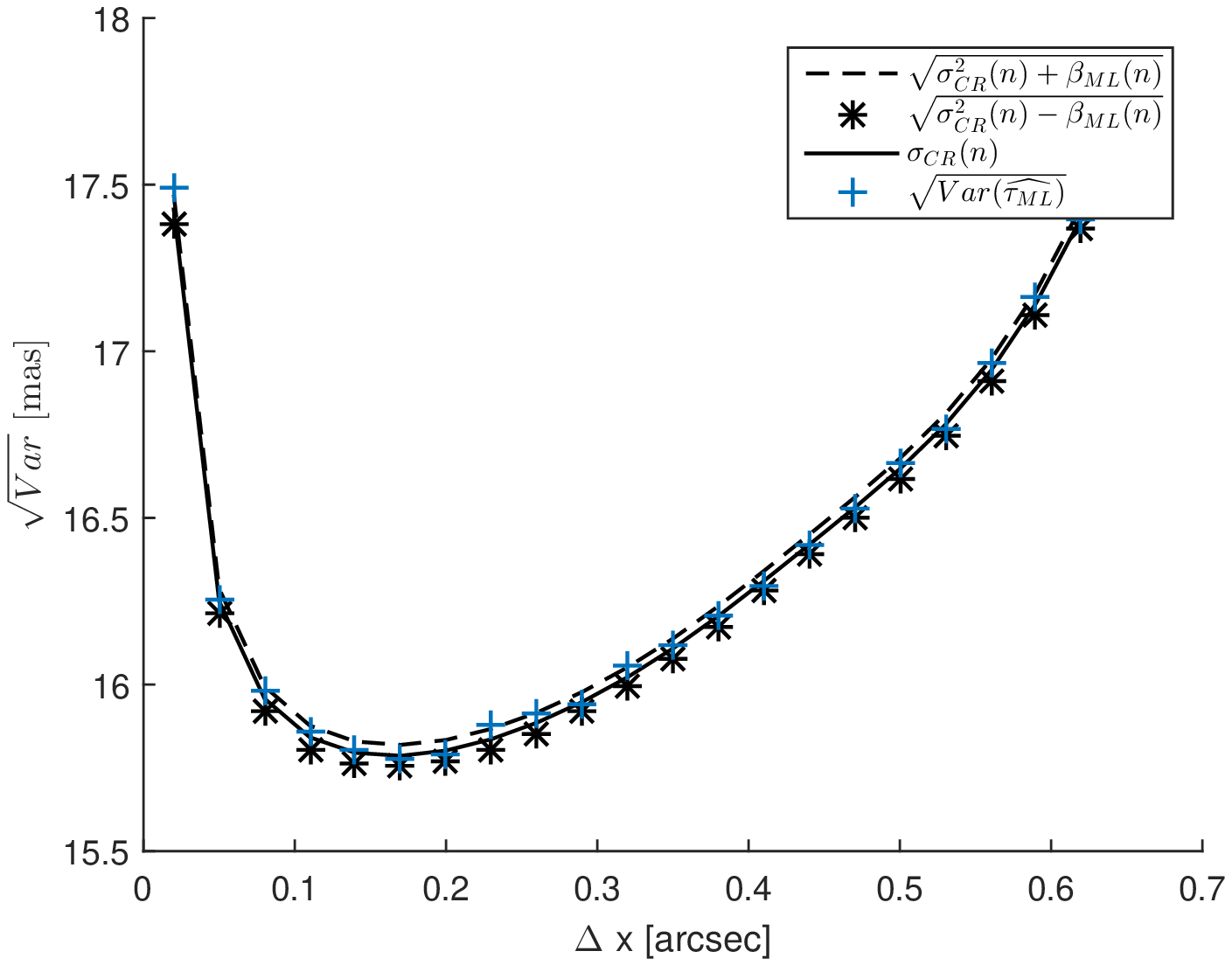} 
      \label{fig2:b} 
  \end{minipage} 
  \begin{minipage}[b]{0.5\linewidth}
    \centering
    \includegraphics[width=1.05\linewidth]{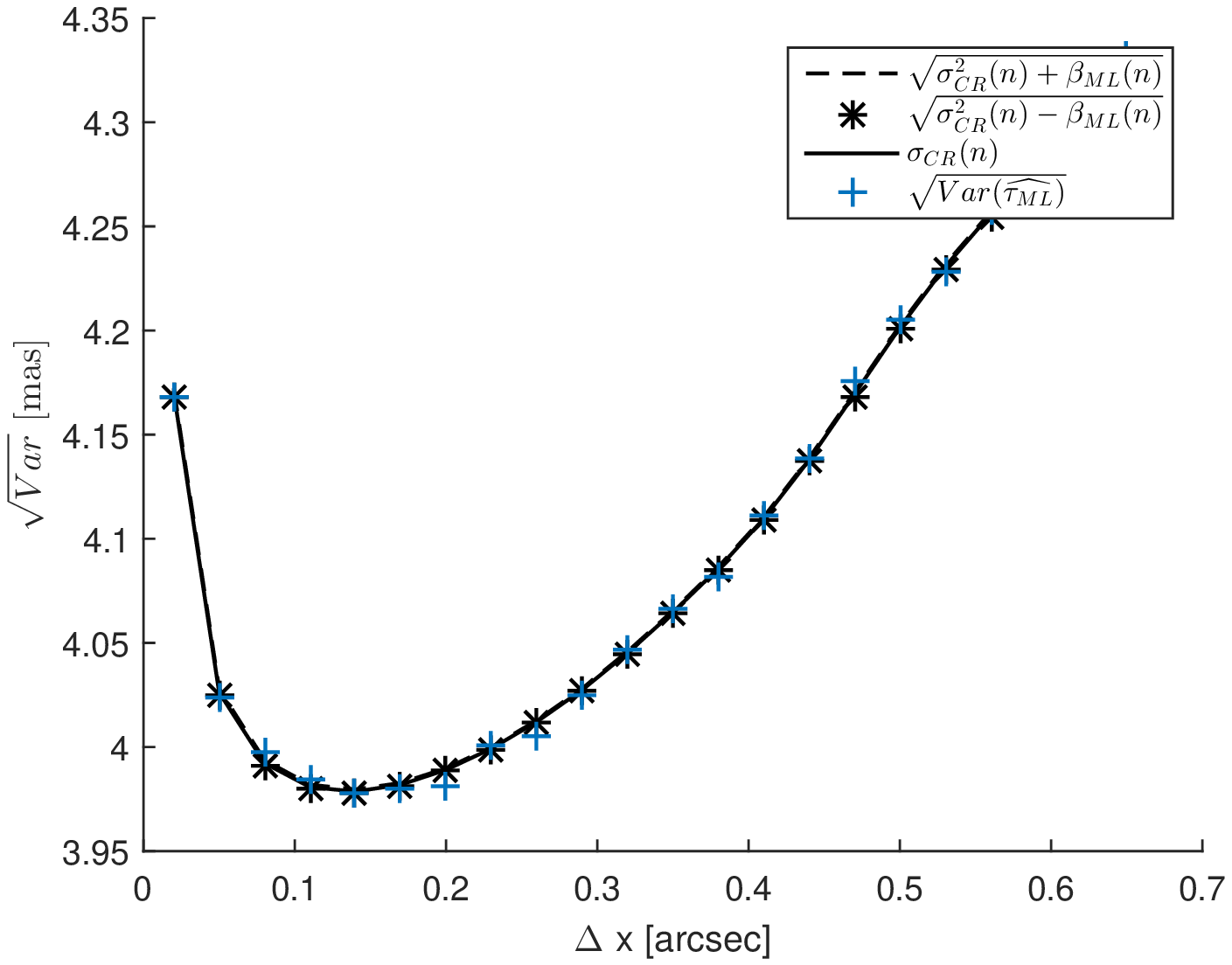} 
      \label{fig2:c}  
  \end{minipage}
  \hfill
  \begin{minipage}[b]{0.5\linewidth}
    \centering
    \includegraphics[width=1.05\linewidth]{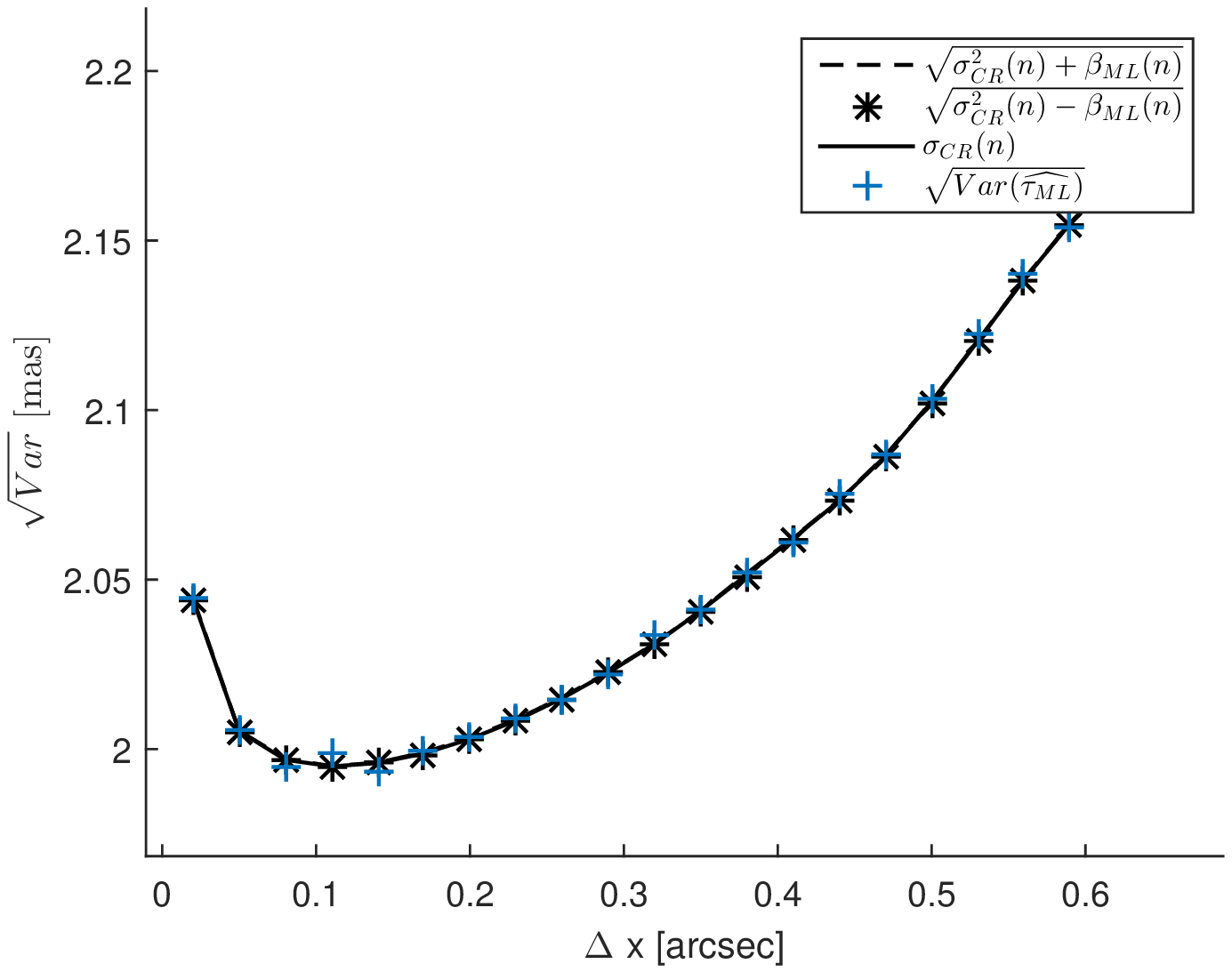} 
      \label{fig2:d} 
  \end{minipage} 
  \caption{Range of the square root of the variance performance (in
    miliarcsecond=mas) for the ML method in astrometry as predicted by
    Theorems 1 and 3,
      Eq.~(\ref{eq_sec_main_astrometry_11}). Results are reported for
    different representative values of $\tilde{F}$ and across
    different pixel sizes (top-left to bottom-right): $\tilde{F} \in
    \left\{1080, 3224,20004,60160\right\}$~e$^-$.}
  \label{ML_v}
\end{figure}

Considering an object located in the center of the array, i.e.,
$x_c=x_o$, the performance curves are presented in Fig.~\ref{ML_v} for
$S/N$ $\in \{12,32,120,230 \}$ and $\Delta x\in [0.1,0.65]$~arcsec.
First, we note that there is a significant difference in the
predictions of our methodology for the ML estimator in comparison with
what we predict in the WLS case. In fact, the results of our approach
are very precise for the determination of the variance of the ML
estimator in all the regimes, from small to high $S/N$, which is
remarkable. More important it is the fact that, from these findings,
we observe that the performance deviation from the MVB
in the worse case (small $S/N$) is very small (see Table~\ref{ML_table}, first row), while for any practical purposes the
variance of the ML estimator achieves the CR limit for all the other
regimes, from medium to high $S/N$, which is a numerical corroboration
of the optimality of the ML estimator in astrometry, as predicted
theoretically by Theorems 1 and 3. In Fig.~\ref{ML_v} we
  also show the square root of the empirical variance
  ($Var(\hat{\tau_{ML}})$) with respect to the empirically-determined
  mean position $\hat x_c$ (using the ML estimator), all as derived
  from the simulations, showing good consistency with our
  predictions.

\begin{figure}[tbp] 
  \begin{minipage}[b]{0.5\linewidth}
  \centering
    \includegraphics[width=1.04\linewidth]{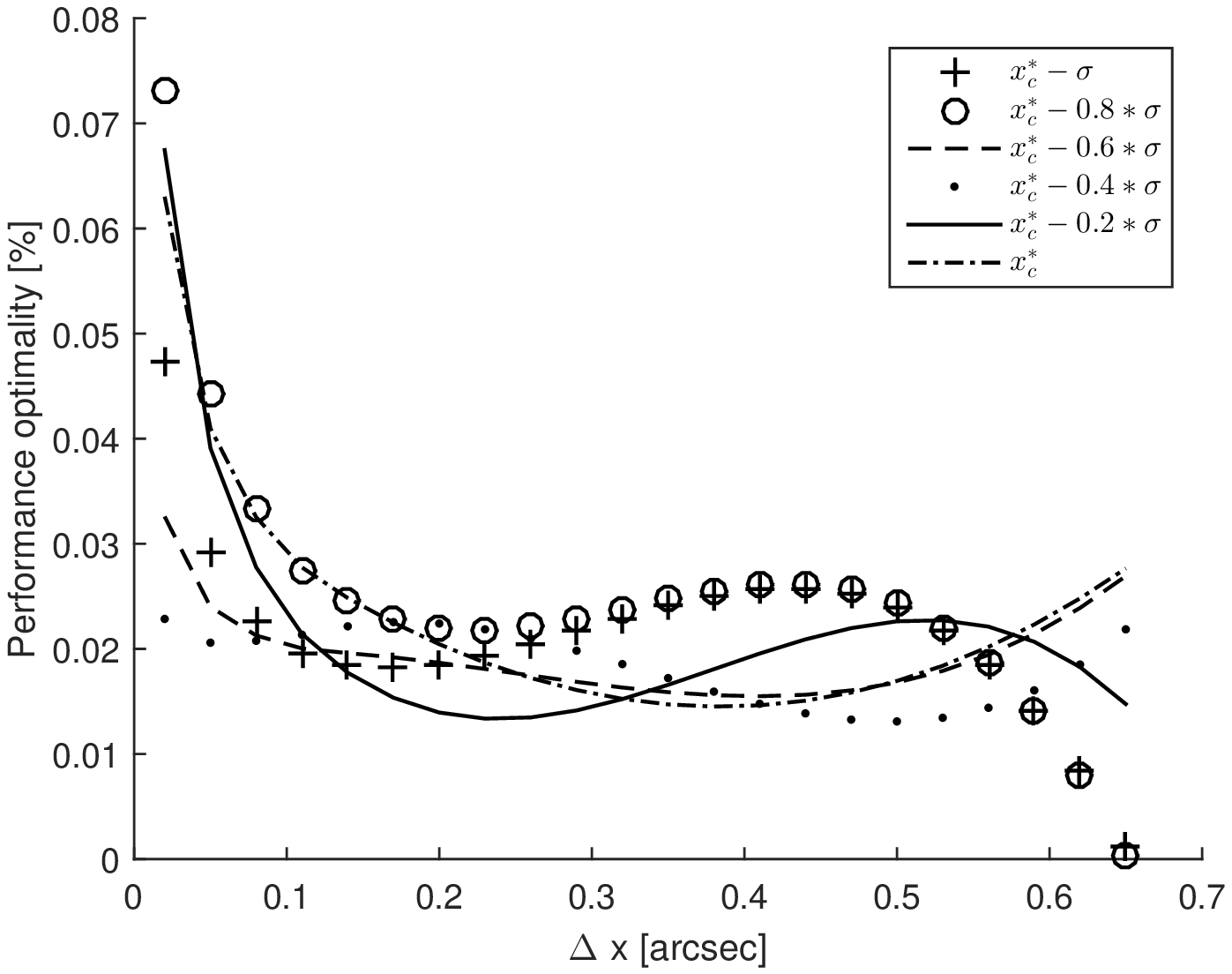} 
  \end{minipage} 
  \begin{minipage}[b]{0.5\linewidth}
    \centering
    \includegraphics[width=1.04\linewidth]{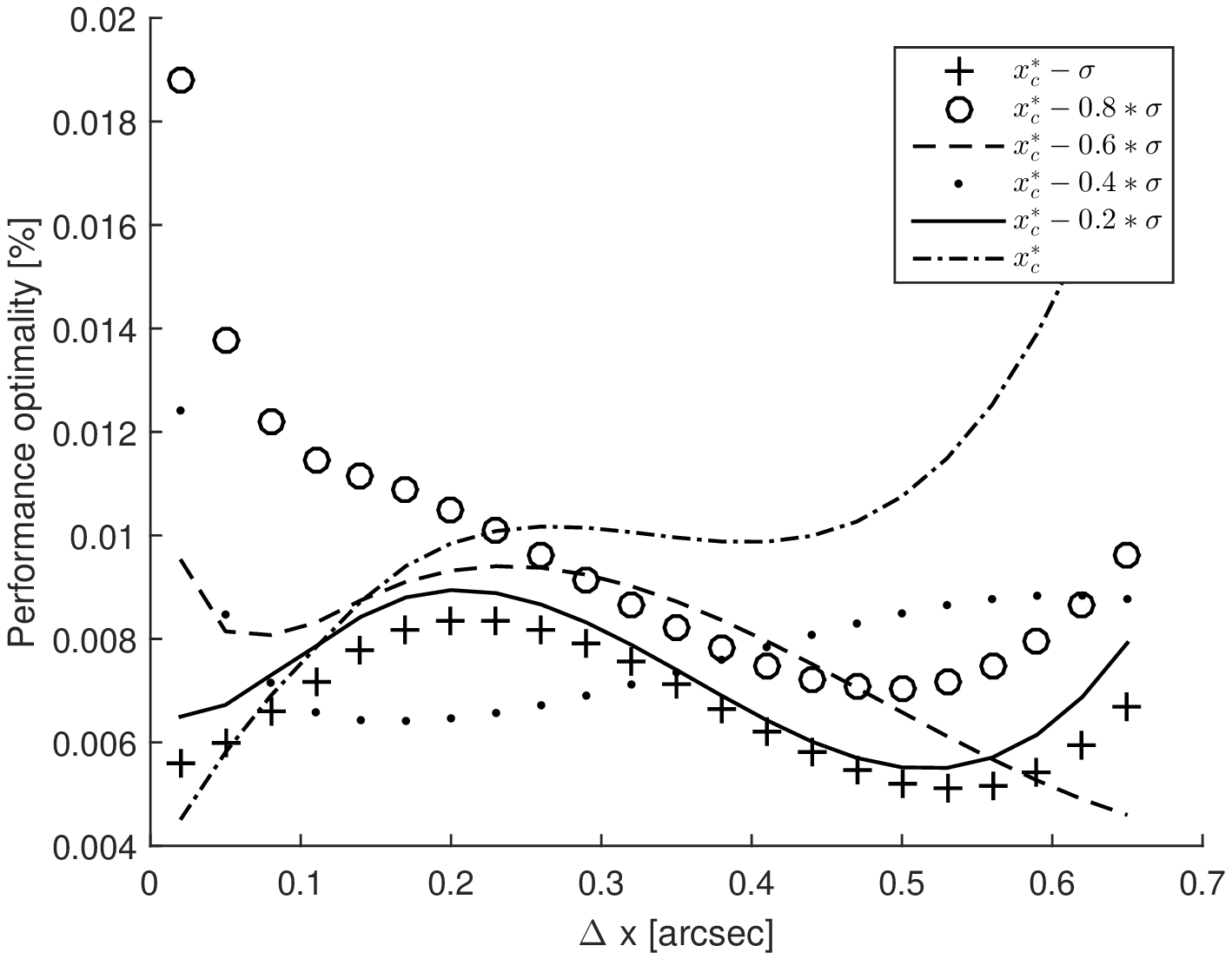} 
  \end{minipage} 
   \label{ML_weight}
   \caption{Performance optimality of the ML
     estimator (computed as $100\times
     \frac{\sqrt{\sigma^2_{ML}(n)+\beta_{ML}
         (n)}-\sigma_{ML}(n)}{\sigma_{ML}(n)}$) for different
     positions of the target object $x_c\in \Theta=
     \{x_o^{\ast}-\sigma , x_o^{\ast}-0.8*\sigma
     ,x_o^{\ast}-0.6*\sigma ,x_o^{\ast}-0.4*\sigma,
     x_o^{\ast}-0.2*\sigma, x_o^{\ast} \}$ in the array, as a function
     of pixel resolution. The left panel shows the case
     $\tilde{F}=20004$~e$^-$, the right panel the case
     $\tilde{F}=60160$~e$^-$.}
    \label{ML_relative_error}
\end{figure}

Complementing this analysis, we conducted the same comparison
considering different positions for the object within the array
obtaining the same trends and conclusions. To summarize these results,
Fig.~\ref{ML_relative_error} shows the value $100\times
\frac{\sqrt{\sigma^2_{ML}(n)+\beta_{ML}
    (n)}-\sigma_{ML}(n)}{\sigma_{ML}(n)}$, which is an indicator of
the quality of the estimator (the smaller the better) for different
scenarios of the position of the object $x_c\in \Theta=
\{x_o^{\ast}-\sigma , x_o^{\ast}-0.8*\sigma ,x_o^{\ast}-0.6*\sigma
,x_o^{\ast}-0.4*\sigma, x_o^{\ast}-0.2*\sigma, x_o^{\ast} \}$. In
particular, for all the evaluated positions, the relative discrepancy
is bounded (relative to the CR bound) by values in the range of
$0.025\%$ and $0.012\%$ for pixel resolution in the range $\Delta x
\in [0.1,0.2]$~arcsec for $S/N=120$ and $S/N=230$,
respectively. Finally, Table~\ref{ML_table} presents the relative
discrepancy for all the range of $S/N$ values considered in this study
for the case $\Delta x=0.2$~arcsec.

\begin{table}[t]
\begin{center}
\begin{tabular}{c|c|c|c|c}
\hline
\hline
Position $x_c$&$S/N$=12 &$S/N$=32&$S/N$=120&$S/N$=230\\
\hline
$x_o$ & 3.8$\%$ & 0.34$\%$ & 0.032$\%$ & 0.010$\%$ \\
\hline
$x_o-0.2*\sigma$ & 4.1$\%$ & 0.27$\%$ & 0.014$\%$ & 0.009$\%$ \\
\hline
$x_o-0.4*\sigma$ & 4.3$\%$ & 0.19$\%$ & 0.022$\%$ & 0.007$\%$ \\
\hline
$x_o-0.6*\sigma$ & 3.8$\%$ & 0.29$\%$ & 0.019$\%$ & 0.009$\%$ \\
\hline
$x_o-0.8*\sigma$ & 3.9$\%$ & 0.30$\%$ & 0.022$\%$ & 0.011$\%$ \\
\hline
$x_o-\sigma$ & 3.6$\%$ & 0.40$\%$ & 0.019$\%$ & 0.008$\%$ \\
\hline
\hline
\end{tabular} 
\caption{Performance quality of the ML estimator relative to the
  \crra\ bound expressed in terms of the indicator $100\times
  \frac{\sqrt{\sigma^2_{ML}(n)+\beta_{ML}
      (n)}-\sigma_{ML}(n)}{\sigma_{ML}(n)}$ from the result in
  Theorem~3.  The results are presented for different positions of the
  object $x_c\in \{x_o^{\ast}-\sigma , x_o^{\ast}-0.8*\sigma
  ,x_o^{\ast}-0.6*\sigma ,x_o^{\ast}-0.4*\sigma,
  x_o^{\ast}-0.2*\sigma, x_o^{\ast} \}$ (rows) and for different $S/N
  \in \left\{12, 32, 120, 230 \right\}$ (columns).}
\label{ML_table}
\end{center}
\end{table}

We conclude from this analysis that the ML estimator in
  nearly optimal for the medium, high and very high $S/N$ regimes
  across pixel resolutions, achieving the MVB for astrometry. This is
  an interesting result, since it lends further support to the
  adoption of these types of estimators for very demanding astrometric
  applications, as has been done in the case of Gaia
  \citep{Linde2008}. We note that \citet{Vaki2016} reach the same
  conclusion regarding the optimality of the ML method in comparison
  with the MVB, through simulations of 2D CCD images using a broad set
  of Moffat PSF stellar profiles. While their results are purely
  empirical, it is interesting that they test the ML using a different
  PSF from ours, and in a 2D scenario, and yet they reach the same
  conclusions. More recently, \citet{Gai2017} have tested (also
  empirically) the ML method in a 1D scenario (similar to ours), but
  in the context of a Gaia-like PSF. They find that the ML is unbiased
  (in agreement with our results, see Fig.~\ref{WLS_bias_relative}),
  and, by comparing two implementations of the ML they conclude that
  they predict self-consistent and reliable results over a broad range
  of flux, background, and instrument response variations. It would
  still be quite interesting to compare the performance of those
  implementations against the CR MVB in order to further test our
  theoretical predictions.

\subsection{Comments on an adaptive WLS estimator for astrometry}
\label{sub_sec_adaptive_WLS}

In Sect.~\ref{sub_sec_wls} we presented results that offer a nominal
prediction for the variance of the WLS method through
Eq.~(\ref{eq_sec_main_astrometry_5}) which turns out to be very
accurate in the regime from medium to high $S/N$ as shown in
Sect.~\ref{subsec_anal_wls}. Interestingly, 
  Remark 2 tells us that this nominal values
precisely match the CR limit for an optimal selection of weights given
in Eq.~(\ref{eq_sec_main_astrometry_7}), namely $w_i \sim
1/\lambda_i(x_c)$ for all $i=1,..,n$ (compare
Eqs.~(\ref{eq_sec_main_astrometry_6}) and
(\ref{eq_sec_main_astrometry_12})). As this selection is unfeasible,
because it requires the knowledge of $x_c$ (see the expression in
Eq.~(\ref{eq_pre_2b})), we can approximate this value by a noisy
version of it, considering the fact that the expected value of the
observations $I_i$ that we measure at pixel $i$ is $\lambda_i(x_c)$
using our model in Eq.~(\ref{eq_pre_4}).  Therefore $I_i$ can be
interpreted as a noisy version of $\lambda_i(x_c)$ and
\begin{equation}\label{eq_sub_sec_adaptive_WLS_0}
\hat{w}_i(I_i)=\frac{1}{I_i}
\end{equation}
can be seen as a noisy version of the ideal weight
$\frac{1}{\lambda_i(x_c)}$.  Adopting this data-driven weighting
approach, we would have an adaptive WLS method as the weights are not
fixed but instead they are a function of the data $\left\{ I_1:
i=1,..,n \right\}$.  This selection of weights can be interpreted as
an empirical version of the optimal weights that achieves the CR
bound.  Then, the problem reduces to solve
\begin{equation}\label{eq_sub_sec_adaptive_WLS_1}
\tau_{AWLS}(I^n) = \arg \min_{\alpha
  \in\mathbb{R}}J_{AWLS}(\alpha,I^n)
\end{equation}
where
\begin{equation}\label{eq_sub_sec_adaptive_WLS_2}
J_{AWLS}(\alpha,I^n)=\sum_{i=1}^n
\hat{w}_i(I_i)(I_i-\lambda_i(\alpha))^2.
\end{equation}

\begin{figure}[tbp]
  \begin{minipage}[b]{0.5\linewidth}
  \centering
    \includegraphics[width=1.05\linewidth]{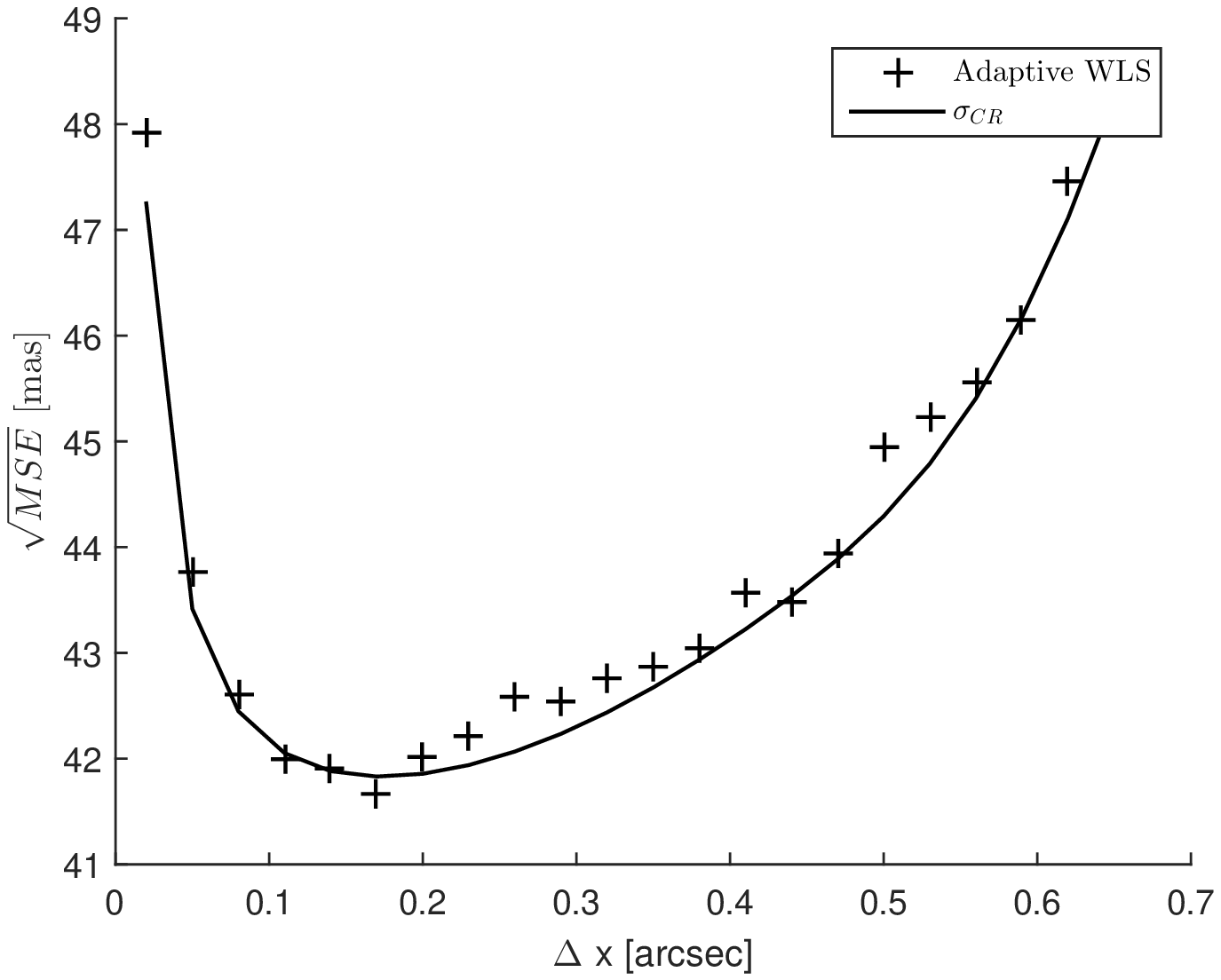} 
  \end{minipage} 
  \begin{minipage}[b]{0.5\linewidth}
    \centering
    \includegraphics[width=1.05\linewidth]{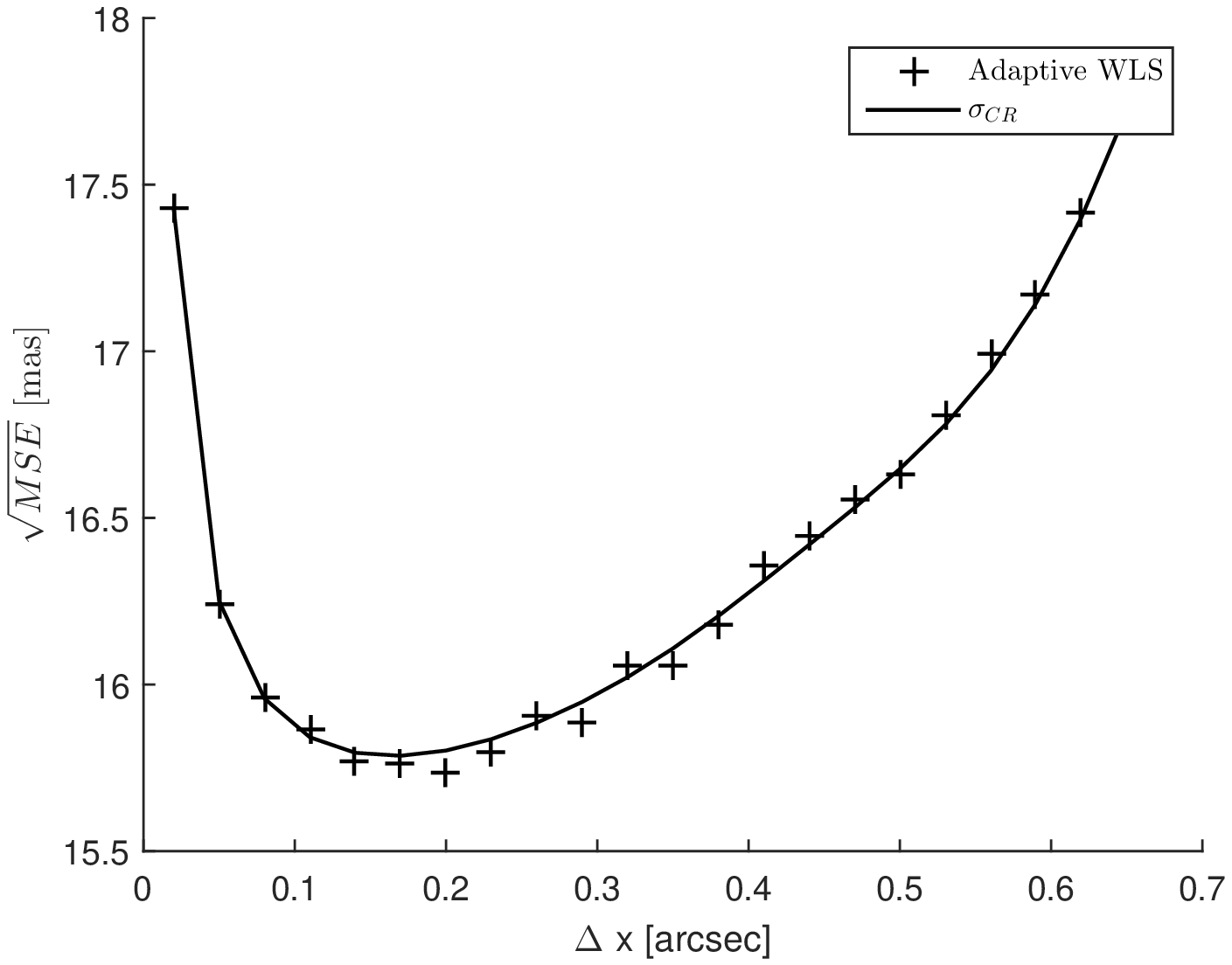}  
  \end{minipage} 
  \begin{minipage}[b]{0.5\linewidth}
    \centering
    \includegraphics[width=1.05\linewidth]{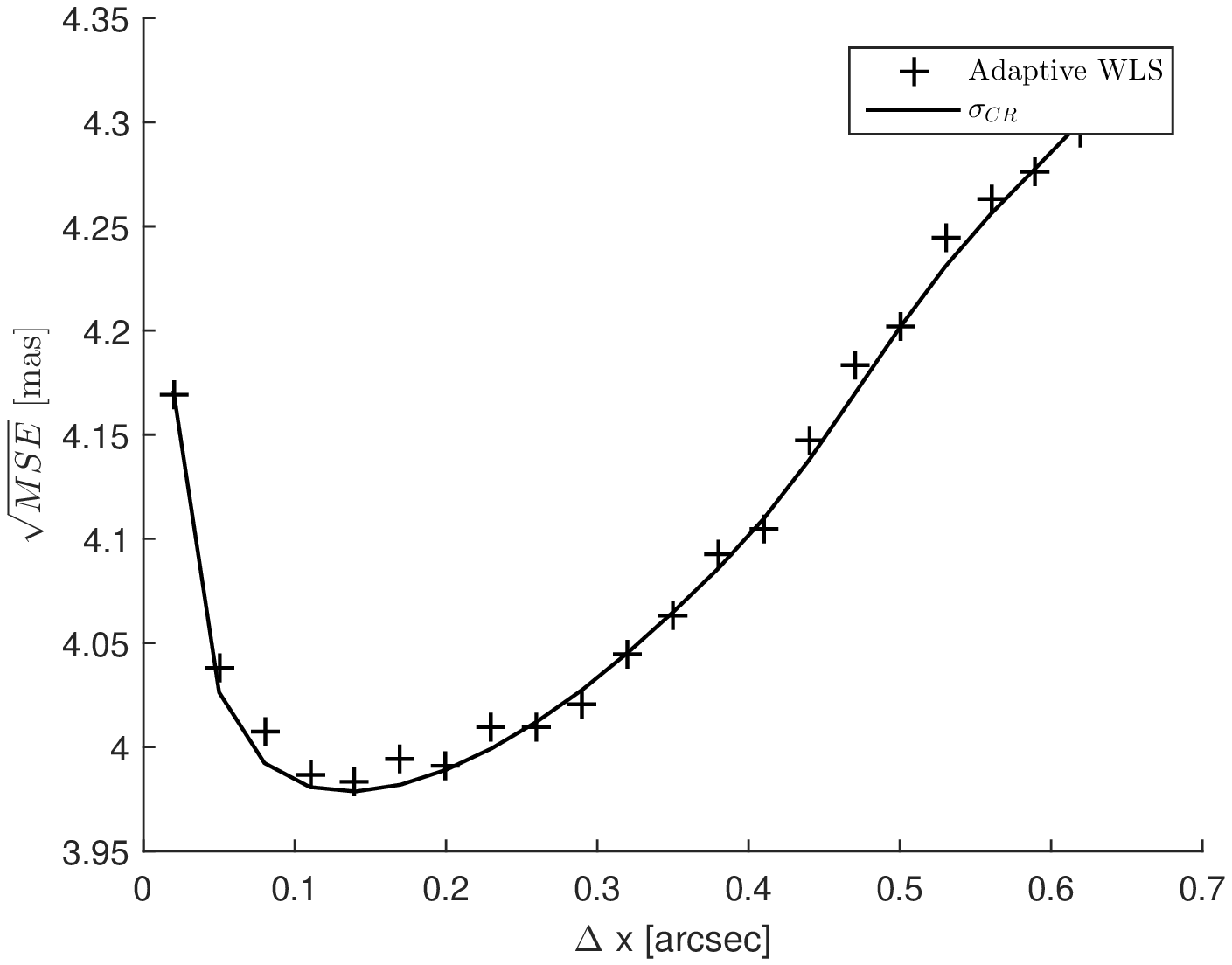}   
  \end{minipage}
  \hfill
  \begin{minipage}[b]{0.5\linewidth}
    \centering
    \includegraphics[width=1.05\linewidth]{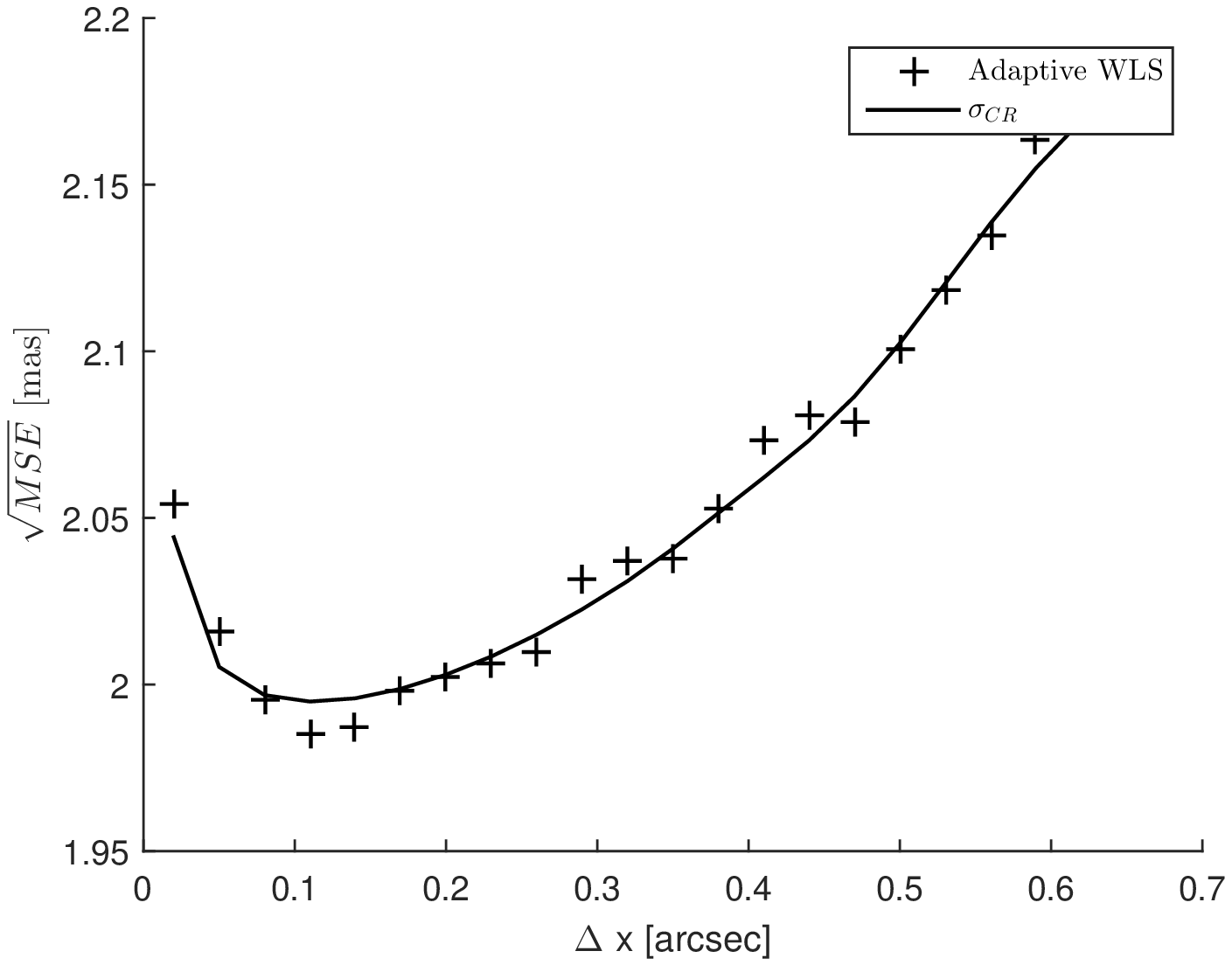}   
  \end{minipage} 
  \caption{Performance comparison between the $\sqrt{MSE}$ of the
    adaptive WLS estimator and $\sigma_{CR}(n)$, both in mas. Results
    are reported for different $\tilde{F}$ and across different pixel
    sizes: (top-left to bottom-right) $\tilde{F} \in \left\{1080,
    3224,20004,60160\right\}$~e$^-$.}
   \label{AWLS}
\end{figure}

Fig.~\ref{AWLS} presents the performance of this scheme for the same
regimes we have been exploring in this work, supporting the conjecture
that this selection of weights resembles the optimal weight selection
and in fact achieves MSE performances that are surprisingly close to
the CR bound in all the observational regimes.

Our results in this sub-section show that some commonly adopted
weighting schemes, using the analogous to
Eq.~(\ref{eq_sub_sec_adaptive_WLS_0}), are a very good data-driven
choice for methods that employ a WLS scheme, such is the case, e.g.,
of the well know PSF stellar fitting program (including astrometry)
DAOPHOT, described in \citet[Eq.~(10)]{stetson1987daophot}.


\newpage
\subsection{Undersampled case}
\label{undersampledcase}
So far we have analyzed the performance of the ML and WLS estimators
in regimes where the PSF (assumed to have a $FWHM=1$~arcsec) is well
sampled by the detector. In this section we analyze the undersampled
case, where we have only a few pixels that capture most the flux of
the source. This scenario would be typical of nearly
diffraction-limited images taken, e.g., from the space, or with an AO
system from the ground.

Table \ref{table_undersampled} shows the performance of the bounds
presented in Theorems~2 and~3 in the undersampled regime, where the
PSF is concentrated in only a few pixels (order of 5 to 9 pixels), and
where the flux is relatively large. For the space-based case, the
background is reduced almost to the purely instrument noise (left
numbers in the table), while in the ground-based case the background
would be larger, but we would still have a very large Strehl
ratio. From the table it can be seen that the performance is nearly
optimal with the ML estimator, even in the severely undersampled
regime (small FWHM values). On the other hand, the WLS deteriorates
its performance very quickly as the image size decreases, reaching up
to nearly 45\% when the FWHM is on the order of the detector pixel
size.

\begin{table}[h]
\begin{center}
\begin{tabular}{|c|c|c|c|c|}
\hline
FWHM &ML error [$\%$] & WLS error [$\%$]\\
\hline
$0.2$ & 0.03 / 0.02 & 25 / 45\\
\hline
$0.3$ & 0.03 / 0.02 & 5.5 / 8.0 \\
\hline
$0.4$ & 0.01 / 0.03 & 1.9 / 2.8 \\
\hline
$0.5$ &  0.01 / 0.04 & 1.4 / 2.3 \\
\hline
\end{tabular} 

\caption{Performance quality of the ML and WLS estimators relative to
  the nominal bound expressed in terms of the indicator $100\times
  \frac{\sqrt{\sigma^2_{ML}(n)+\beta_{ML}
      (n)}-\sigma_{ML}(n)}{\sigma_{ML}(n)}$ and $100\times
  \frac{\sqrt{\sigma^2_{WLS}(n)+\beta_{WLS}
      (n)}-\sigma_{WLS}(n)}{\sigma_{WLS}(n)}$, respectively in the
  undersampled regime (space based and ground based are on the left
  and right, respectively).  The results are presented for different
  values of $FWHM$ (measured in arcsec) with $\tilde{F}=20004 e^-$, $\tilde{B}=25 e^-$ (space based), $\tilde{B}=626 e^-$ (ground based) and $\Delta x=0.2$ arcsec.}
\label{table_undersampled}
\end{center}
\end{table}

We emphasize that the above results are computed in the ideal
situation of a uniform pixel response function (and a perfect
flat-field correction), whereas in reality a non-uniform pixel
response has a critical impact on the \crra\ bound as demonstrated by
\citet[Figure~1]{adorf1996}. In addition, in the case of stare-mode
space-based images acquired under very small background, charge
transfer inefficiency effects due to time-varying charge-traps in the
detector would also blur (and systematically shift) the images,
deteriorating the ultimate astrometric precision above the \crra\ limit
(for details, see, e.g., \citet{bristet06}, especially their Figure
10).

\section{Summary, conclusions, and outlook}
\label{final}

In this paper we study the performance of the WLS and ML estimators
for relative astrometry on digital detectors subject to Poisson noise,
in comparison with the best possible attainable precision given by the
CR bound. Our study includes analytical results, and numerical
simulations under realistic observational conditions to help us
to corroborate our theoretical findings.

By generalizing the proposal by \citet{fessler1996} we are able to
obtain, for the first time, close-form expressions for the variance
and the mean of implicit estimators (as is the particular case of the
WLS and ML schemes), which can be computed directly from the data
(see Theorem~1, in particular Eqs.~(\ref{bias_eqn})
and~(\ref{VAR_eqn}), and Appendix~A). When specifying this result to
astrometry with digital detectors, we are able to bound both the bias
and the variance of the relative position of a celestial source on a
CCD array as a function of all the relevant parameters of the problem
(see Eqs.~(\ref{eq_sec_main_astrometry_4})
and~(\ref{eq_sec_main_astrometry_5}) or
Eqs.~(\ref{eq_sec_main_astrometry_10}) and
~(\ref{eq_sec_main_astrometry_11}) for the WLS and ML estimators
respectively). We verified that the bias of the WLS and ML methods are
negligible in all the observational regimes explored in this paper
(see Fig.~\ref{WLS_bias_relative}).

A careful analysis of our predictions confirms earlier results by
\citet{2015lobos} (for the LS method) in that the WLS method is, in
general, sub-optimal (in comparison with the MVB given by the CR
result), specially at high and very high $S/N$ (see the two bottom
panels on Fig.~\ref{WLS_v}). However, a judicious data-driven
selection of weights (called ``adaptive'' WLS method by us,
Sect.~(\ref{sub_sec_adaptive_WLS})), improves the performance of the
WLS substantially (see Fig.~\ref{AWLS}). This is an interesting
result, given the widespread use and simple numerical implementation
of the WLS method.

The ML method is found to have both a smaller bias than the WLS method
(compare left and right panels of Fig.~\ref{WLS_bias_relative}),
although the bias on both methods is already quite small, and a tight
correspondence to the MVB throughout the entire range of $S/N$ regimes
explored in this paper (Fig.~\ref{ML_v}). Therefore, the ML
estimator for astrometry is consistently optimal, and should be the
estimator of choice for high-precision applications.

This paper, along with \citet{2015lobos}, completes an in-depth study
of the performance of commonly used estimators in astrometry using
PIDs, and sets the stage for the development of codes that could
efficiently implement astrometric ML estimators on 2D detectors,
incorporating also the simultaneous measurement of fluxes, as explored by \citet{Gai2017}.
\section{Acknowledgments}

SE acknowledges support from CONICYT-PCHA/MagisterNacional/2016 -
22160840. The authors want to thank the support of the Advanced Center
for Electrical and Electronic Engineering, AC3E, Basal Project FB0008,
PIA ACT1405, and the Chilean Centro de Excelencia en Astrofisica y
Tecnologias Afines (CATA) BASAL PFB/06 from CONICYT. The authors also
acknowledge support by CONICYT/FONDECYT grants \# 1151213, 1170044,
and 1170854. RAM also acknowledges support from the Project IC120009
Millennium Institute of Astrophysics (MAS) of the Iniciativa
Cientifica Milenio del Ministerio de Economia, Fomento y Turismo de
Chile. Finally, we thank an anonymous referee for his/her careful reading of our manuscript, and for pointing out the exploration of  the relevant space-based undersampled regime that led to the discussion in Section \ref{undersampledcase}.

\begin{appendix}
\section{Proof of Theorem 1}
\label{proof_th_main_general}
We begin presenting the expressions for $(\epsilon_{J} (n), \beta_{J} (n),\sigma^2_{J} (n))$ to complete 
the statement of the result.  
\begin{equation}\label{eq_proof_th_main_general_1}
\epsilon_{J} (n) = \max_{t \in [0,1]}\left \rvert \mathbb{E}_{I^n \sim f_{x_c}} \left \{ \frac{1}{2}\sum_{i=1}^n\sum_{j=1}^n\frac{\partial^2}{\partial I_i\partial I_j}\tau_J(\bar{I}^n-t(I^n-\bar{I}^n))(I_i-\bar{I}_i)(I_j-\bar{I}_j) \right \}\right\rvert, 
\end{equation}
\begin{equation}\label{eq_proof_th_main_general_2}
\beta_{J} (n) =\epsilon'_{J} (n) + 2\delta'_{J}(n), 
\end{equation}
where 
\begin{equation}\label{eq_proof_th_main_general_3}
\epsilon'_{J} (n) =\max_{t \in [0,1]}\mathbb{E}_{I^n \sim f_{x_c}} \left \{ \left ( \frac{1}{2}\sum_{i=1}^n\sum_{j=1}^n\frac{\partial^2}{\partial I_i\partial I_j}\tau_J(\bar{I}^n-t(I^n-\bar{I}^n))(I_i-\bar{I}_i)(I_j-\bar{I}_j)  \right )^2 \right \},
\end{equation}
\begin{equation} \label{eq_proof_th_main_general_4}
\delta'_{J} (n)=\max_{t \in [0,1]}\left \rvert \mathbb{E}_{I^n \sim f_{x_c}} \left \{ (\nabla \tau_J(\bar{I}^n) \cdot (I^n-\bar{I}^n))\cdot \frac{1}{2}\sum_{i=1}^n\sum_{j=1}^n\frac{\partial^2}{\partial I_i\partial I_j}\tau_J(\bar{I}^n-t(I^n-\bar{I}^n))(I_i-\bar{I}_i)(I_j-\bar{I}_j) \right \}\right\rvert,
\end{equation}
and, finally, 
\begin{equation} \label{eq_proof_th_main_general_5} 
\sigma^2_{J} (n) = \left [\frac{\partial^2J(\tau_J(\bar{I}^n),\bar{I}^n)}{\partial \alpha^2}\right ]^{-1}\left [\frac{\partial^2J(\tau_J(\bar{I}^n),\bar{I}^n)}{\partial \alpha\partial I_i}\right ]Cov\{I^n\}\left [\frac{\partial^2J(\tau_J(\bar{I}^n),\bar{I}^n)}{\partial \alpha\partial I_i}\right ]^T\left [\frac{\partial^2J(\tau_J(\bar{I}^n),\bar{I}^n)}{\partial \alpha^2}\right ]^{-1}.
\end{equation}

\textit{Proof of Theorem 1:}
Using the chain rule in the cost function $J(\alpha,I^n)$ and taking the partial derivative $\frac{\partial}{\partial I_i}$ of both sides in (\ref{eq_main_general_2}), we have that
\begin{equation}\label{eq_null_der_sys}
0=\frac{\partial^2}{\partial\alpha^2}J(\tau(I^n),I^n)\frac{\partial}{\partial I_i}\tau(I^n)+\frac{\partial^2}{\partial\alpha\partial I_i}J(\tau(I^n),I^n),~ i=1,\ldots,n.
\end{equation}
Thus, we have $n$ equations with one unknown, and it holds for any $I^n$. Defining the operators
\begin{eqnarray}
\nabla^{20}(\cdot )=\frac{\partial^2}{\partial \alpha^2}, 
\nabla^{11}(\cdot )=\frac{\partial^2}{\partial \alpha\partial I_i},
\end{eqnarray}
of dimensions $1\times 1$ and $1\times n$, respectively, we can  express (\ref{eq_null_der_sys}) in matrix form as
\begin{equation}\label{eq_null_der_matrix}
0=\nabla^{20}J(\tau(I^n),I^n)\nabla\tau(I^n)+\nabla^{11}J(\tau(I^n),I^n).
\end{equation}
Assuming that the matrix $\nabla^{20}J(\tau(I^n),I^n)$ is non singular,  we can calculate $\nabla\tau(I^n)$ from (\ref{eq_null_der_matrix})
\begin{equation}\label{eq_nabla_theta}
\nabla\tau(I^n)=-[\nabla^{20}J(\tau(I^n),I^n)]^{-1}\nabla^{11}J(\tau(I^n),I^n).
\end{equation}
Finally, using (\ref{eq_nabla_theta}), evaluating at $\bar{I}^n$, and then replacing in (\ref{eq_main_general_5}), we have that
\begin{eqnarray}\label{sigma}
&\sigma^2_{J}(n)& = -[\nabla^{20}J(\tau(\bar{I}^n),\bar{I}^n)]^{-1}\nabla^{11}J(\tau(\bar{I}^n),\bar{I}^n)Cov\{I^n\}(-[\nabla^{20}J(\tau(\bar{I}^n),\bar{I}^n)]^{-1}\nabla^{11}J(\tau(\bar{I}^n,\bar{I}^n))^T\nonumber\\
&=&[\nabla^{20}J(\tau(\bar{I}^n),\bar{I}^n)]^{-1}\nabla^{11}J(\tau(\bar{I}^n),\bar{I}^n)Cov\{I^n\}[\nabla^{11}J(\tau(\bar{I}^n),\bar{I}^n))]^T[\nabla^{20}J(\tau(\bar{I}^n),\bar{I}^n)]^{-1}.\nonumber\\
&=& \left [\frac{\partial^2J(\tau(\bar{I}^n),\bar{I}^n)}{\partial \alpha^2}\right ]^{-1}\left [\frac{\partial^2J(\tau(\bar{I}^n),\bar{I}^n)}{\partial \alpha\partial I_i}\right ]Cov\{I^n\}\left [\frac{\partial^2J(\tau(\bar{I}^n),\bar{I}^n)}{\partial \alpha\partial I_i}\right ]^T\left [\frac{\partial^2J(\tau(\bar{I}^n),\bar{I}^n)}{\partial \alpha^2}\right ]^{-1}.
\end{eqnarray}
Moving into the residual term $\gamma_{J} (n)$ in (\ref{eq_main_general_5}) captured by $\beta_{J}(n)$, we must consider the variance of the error function $Var\{ e(\bar{I}^n,I^n-\bar{I}^n) \}$ and the covariance $Cov\{\nabla\tau_J(\bar{I}^n) (I^n-\bar{I}^n),e(\bar{I}^n,I^n-\bar{I}^n)\}$. For the first, we have that 
\begin{eqnarray}
Var\{ e(\bar{I}^n,I^n-\bar{I}^n)\}&= & Var\left \{\frac{1}{2}\sum_{i=1}^n\sum_{j=1}^n\frac{\partial^2\tau}{\partial I_i \partial I_j} (\bar{I}^n+t(I^n-\bar{I}^n))(I_i-\bar{I_i})(I_j-\bar{I_j})\right \}\nonumber\\
&\leq & \mathbb{E}\left \{\left (\frac{1}{2}\sum_{i=1}^n\sum_{j=1}^n\frac{\partial^2\tau}{\partial I_i \partial I_j} (\bar{I}^n+t(I^n-\bar{I}^n))(I_i-\bar{I_i})(I_j-\bar{I_j}) \right )^2\right \} \nonumber\\
&\leq & \underbrace{\max_{t \in [0,1]}\mathbb{E}\left \{\left (\frac{1}{2}\sum_{i=1}^n\sum_{j=1}^n\frac{\partial^2\tau}{\partial I_i \partial I_j} (\bar{I}^n+t(I^n-\bar{I}^n))(I_i-\bar{I_i})(I_j-\bar{I_j}) \right )^2\right \}}_{ = \epsilon'_{J}(n)}
\label{Var_bound}
\end{eqnarray}
On the other hand, for the covariance,  using the main assumption in (\ref{close_gradient}), it is clear that
\begin{equation}
\mathbb{E}_{I^n \sim f_{x_c}}\left \{\nabla \tau_{ML}(\bar{I}^n) \cdot (I^n-\bar{I}^n)\right \} =\mathbb{E}_{I^n \sim f_{x_c}}\left \{ a\sum_{i=1}^n\limits b_i(I_i-\bar{I}_i)\right \} =0.
\end{equation}
From this,
\begin{eqnarray}
&&|Cov\{\nabla \tau(\bar{I}^n) (I^n-\bar{I}^n),e(\bar{I}^n,I^n-\bar{I}^n)\}| \nonumber\\
&=&  |\mathbb{E}\left \{  \nabla \tau(\bar{I}^n) (I^n-\bar{I}^n) \left ( e(\bar{I}^n,I^n-\bar{I}^n)-\mathbb{E} \left ( e(\bar{I}^n,I^n-\bar{I}^n)\right )\right ) \right \}| \nonumber\\
&=& \left \rvert \mathbb{E}_{I^n \sim f_{x_c}} \left \{ (\nabla \tau(\bar{I}^n) \cdot (I^n-\bar{I}^n))\cdot \frac{1}{2}\sum_{i=1}^n\sum_{j=1}^n\frac{\partial^2}{\partial I_i\partial I_j}\tau(\bar{I}^n-t(I^n-\bar{I}^n))(I_i-\bar{I}_i)(I_j-\bar{I}_j) \right \}\right\rvert \nonumber\\
&\leq & \underbrace{ \max_{t \in [0,1]}  \left \rvert \mathbb{E}_{I^n \sim f_{x_c}} \left \{ (\nabla \tau(\bar{I}^n) \cdot (I^n-\bar{I}^n))\cdot \frac{1}{2}\sum_{i=1}^n\sum_{j=1}^n\frac{\partial^2}{\partial I_i\partial I_j}\tau(\bar{I}^n-t(I^n-\bar{I}^n))(I_i-\bar{I}_i)(I_j-\bar{I}_j) \right \}\right\rvert}_{= \delta'_{J}(n)}
     \label{cov_bound}
\end{eqnarray}
Finally, replacing (\ref{Var_bound}) and (\ref{cov_bound}) in the definition of $\gamma_{J}(n)$, we have that:
\begin{eqnarray}
|\gamma_{J}(n) |&\leq & Var\{ e(\bar{I}^n,I^n-\bar{I}^n) \}+2|Cov\{\nabla \tau(\bar{I}^n) (I^n-\bar{I}^n),e(\bar{I}^n,I^n-\bar{I}^n)\}| \nonumber \\
&\leq &\epsilon '_{J} (n)+ 2\delta'_{J}(n) = \beta_{J}(n).
\end{eqnarray}
For the bias expression of the result in (\ref{bias_eqn}), using the hypothesis in (\ref{noise_free}), we can take expectation at both sides of (\ref{eq_main_general_4}) to obtain that
\begin{equation}
\begin{split}
|\mathbb{E}_{I^n \sim f_{x_c}}\{ \tau(I^n)\}-x_c |&= \left \rvert \mathbb{E}_{I^n \sim f_{x_c}} \left \{ a\sum_{i=1}^N\limits b_i(I_i-\bar{I}_i)+ e(\bar{I}^n,I^n-\bar{I}^n)\right \}\right\rvert \\
&= \left \rvert \mathbb{E}_{I^n \sim f_{x_c}} \left \{ e(\bar{I}^n,I^n-\bar{I}^n)\right \}\right\rvert \\
&= \left \rvert \mathbb{E}_{I^n \sim f_{x_c}} \left \{ \frac{1}{2}\sum_{i=1}^n\sum_{j=1}^n\frac{\partial^2}{\partial I_i\partial I_j}\tau(\bar{I}^n-t(I^n-\bar{I}^n))(I_i-\bar{I}_i)(I_j-\bar{I}_j) \right \}\right\rvert \\
&\leq \underbrace{\max_{t \in [0,1]}\left \rvert \mathbb{E}_{I^n \sim f_{x_c}} \left \{ \frac{1}{2}\sum_{i=1}^n\sum_{j=1}^n\frac{\partial^2}{\partial I_i\partial I_j}\tau(\bar{I}^n-t(I^n-\bar{I}^n))(I_i-\bar{I}_i)(I_j-\bar{I}_j) \right \}\right\rvert}_{=\epsilon_{J}(n)}.
\end{split}
\end{equation}

\section{Proof of Theorem 2}
\label{proof_th_wls}
\textit{Proof:} The proof and, in particular, the derivation of
$\sigma^2_{WLS}(n)$, $\beta_{WLS}(n)$ and $\epsilon_{WLS}(n)$ simply reduces to an straightforward application of Theorem 1. For that we need to first
validate the assumptions of Theorem 1.  If we begin
with Eq.~(\ref{eq_nabla_theta})
\begin{equation}\label{eq_proof_th_wls_1}
\nabla\tau(I^n)=-[\nabla^{20}J(\tau_J(I^n),I^n)]^{-1}\nabla^{11}J(\tau_J(I^n),I^n),
\end{equation}
and then we calculate the gradient terms in the RHS of (\ref{eq_proof_th_wls_1}) for our WLS context, it follows that
\begin{eqnarray}\label{eq_proof_th_wls_2}
\nabla^{20}J_{WLS}(\alpha,I^n)&=&\frac{\partial^2}{\partial \alpha^2}J_{WLS}(\alpha,I^n)\nonumber\\
&=& 2\sum_{i=1}^nw_i\left(\left(\frac{\partial \lambda_i(\alpha)}{\partial \alpha}\right)^2+(\lambda_i(\alpha)-I_i)\frac{\partial^2\lambda_i(\alpha)}{\partial \alpha^2}\right),\label{eq_nabla20_astro_WLS}\\
\nabla^{11}J_{WLS}(\alpha,I^n)&=& \left(\frac{\partial^2}{\partial \alpha \partial I_1}J_{WLS}(\alpha,I^n),\ldots,\frac{\partial^2}{\partial \alpha\partial I_n}J_{WLS}(\alpha,I^n)\right)^ T\\
&=&-2\left( w_1\frac{\partial \lambda_1(\alpha)}{\partial \alpha},\ldots, w_n\frac{\partial \lambda_n(\alpha)}{\partial x_c}\right)^ T.\label{eq_nabla11_astro_WLS}
\end{eqnarray}
Following (\ref{eq_proof_th_wls_1}), we need to evaluate
Eqs.~(\ref{eq_nabla20_astro_WLS}) and (\ref{eq_nabla11_astro_WLS}) at
$\alpha=\tau_{WLS}(\bar{I}^n)$. For that, we have the following\footnote{considering that $\bar{I}_i=E\{I_i\}=\lambda_i(x_c)$.}
\begin{equation}\label{eq_proof_th_wls_3}
\tau_{WLS}(\bar{I}^n)= \argmin_{\alpha\in\mathbb{R}} \sum_{i=1}^nw_i(\lambda_i(x_c)-\lambda_i(\alpha))^2. 
\end{equation}
Then we will use the following result:
\begin{proposition}\label{prop_WLS}
Under the assumption of a Gaussian PSF,
$\tau_{WLS}(\bar{I}^n)=x_c$.
\end{proposition}
Notice that this proposition is the second assumption used in Theorem 1. Using this proposition,  we obtain that 
\begin{eqnarray}\label{eq_proof_th_wls_4}
\nabla^{20}J_{WLS}(\tau_{WLS}(\bar{I}^n),\bar{I}^n)
&=& 2\sum_{i=1}^nw_i\left(\left.\left(\frac{\partial \lambda_i(\alpha)}{\partial \alpha}\right)^2\right|_{\alpha=x_c}+(\lambda_i(x_c)-\lambda_i(x_c))\left.\frac{\partial^2\lambda_i(\alpha)}{\partial \alpha^2}\right|_{\alpha=x_c}\right),\nonumber\\
&=& \left.2\sum_{i=1}^nw_i\left(\frac{\partial \lambda_i(\alpha)}{\partial \alpha}\right)^2\right|_{\alpha=x_c},\label{eq_nabla20_astro2_WLS}\\
\nabla^{11}J_{WLS}(\tau_{WLS}(\bar{I}^n),\bar{I}^n)&=&-2\left.\left(w_1\frac{\partial \lambda_1(\alpha)}{\partial \alpha},\ldots,w_n\frac{\partial \lambda_n(\alpha)}{\partial \alpha}\right)^T\right|_{\alpha=\tau_{WLS}(\bar{I}^n)},\nonumber\\
&=&-2\left.\left(w_1\frac{\partial \lambda_1(\alpha)}{\partial \alpha},\ldots,w_n\frac{\partial \lambda_n(\alpha)}{\partial \alpha}\right)^T\right|_{\alpha=x_c}.\label{eq_nabla11_astro2_WLS}
\end{eqnarray}
Finally, applying  (\ref{eq_nabla20_astro2_WLS}) and (\ref{eq_nabla11_astro2_WLS}) in (\ref{eq_nabla_theta}) we have that
\begin{eqnarray}\label{eq_proof_th_wls_5}
 \nabla \tau_{WLS}(\bar{I}^n)\cdot (I^n-\bar{I}^n)&=& -[\nabla^{20}J(\tau_{WLS}(\bar{I}^n),\bar{I}^n)]^{-1} [\nabla^{11}J(\tau_{WLS}(\bar{I}^n),\bar{I}^n)](I^n-\bar{I}^n) \nonumber \\
  &=& \underbrace{\left[\sum_{i=1}^nw_i\left. \left( \frac{\partial\lambda_i(\alpha)}{\partial \alpha}\right)^2\right|_{\alpha=x_c}\right ]^{-1}}_{a}\cdot \sum_{j=1}^n \underbrace{w_j\left.\frac{\partial\lambda_j(\alpha)}{\partial \alpha}\right|_{\alpha=x_c}}_{b_j} (I_j-\mathds{E}(I_j)),
\end{eqnarray}
which offers the decomposition needed for the application of Theorem
1 (Eq.~(\ref{close_gradient})).
For the value of $\sigma^2_{WLS}(n)$ in (\ref{eq_proof_th_main_general_5}), since the observations are independent and follow a Poisson distribution, we have that
\begin{equation} 
\label{eq_cov_astro_ls}
Cov\{I_i,I_j\}=\left\{
  \begin{array}{l l}
    Var\{I_i\}=\lambda_i(x_c), & \quad \text{if $i=j$},\\
    0 & \quad \sim.
  \end{array} \right.
\end{equation}
Then if we replace (\ref{eq_nabla20_astro2_WLS}),
(\ref{eq_nabla11_astro2_WLS}) and (\ref{eq_cov_astro_ls}) in
(\ref{eq_proof_th_main_general_5}),
we have that
\begin{eqnarray}\label{eq_proof_th_wls_7}
\sigma^2_{WLS}(n)&=& \frac{\sum_{i=1}^nw_i^2\lambda_i(x_c)\left.\left(\frac{\partial \lambda_i(\alpha)}{\partial \alpha}\right)^2\right|_{\alpha=x_c}}{\left(\sum_{i=1}^nw_i\left.\left(\frac{\partial \lambda_i(\alpha)}{\partial \alpha}\right)^2\right|_{\alpha=x_c}\right)^2}.
\end{eqnarray}
On the other hand, the expression for $\beta_{WLS}(n)$ and $\epsilon_{WLS}(n)$ can be determined from the evaluation of (\ref{eq_proof_th_main_general_2}) and (\ref{eq_proof_th_main_general_1}), respectively. Looking at them, the problem reduces to determine the key term  $\frac{\partial^2\tau_{WLS}}{\partial I_i \partial I_j} (\bar{I}^n+t(I^n-\bar{I}^n))$. For that,  if we use \cite[Eq.~(17)]{fessler1996} we can obtain the following identity\footnote{The derivation of this identity is presented in Appendix \ref{proof_eq_proof_th_wls_8}.}
\begin{equation}\label{eq_proof_th_wls_8}
\begin{split}
\frac{\partial^2\tau_{WLS}}{\partial I_i \partial I_j} (\bar{I}^n+t(I^n-\bar{I}^n))&=
\frac{-1}{\left [ \sum_{i=1}^n \limits \frac{\partial^2 \lambda_i(\alpha)}{\partial \alpha ^2} \cdot (\lambda_i(\alpha)-(\bar{I}_i+t(I_i-\bar{I}_i)))2w_i+ 2w_i\left ( \frac{\partial \lambda_i(\alpha)}{\partial \alpha }  \right )^2 \right] ^2  }\cdot \\
& \left [ \left [ \left [\sum_{i=1}^n \frac{\partial^3 \lambda_i(\alpha)}{\partial \alpha ^3} \cdot  (\lambda_i(\alpha)-(\bar{I}_i+t(I_i-\bar{I}_i)))2w_i +6w_i\frac{\partial^2 \lambda_i(\alpha)}{\partial \alpha ^2}\frac{\partial \lambda_i(\alpha)}{\partial \alpha }  \right ] \cdot  \right.  \right. \\ 
&  \left. \left.\frac{\left (2w_j\frac{\partial \lambda_j(\alpha)}{\partial \alpha }\right )}{\left [ \sum_{i=1}^n \limits \frac{\partial^2 \lambda_i(\alpha)}{\partial \alpha ^2} \cdot (\lambda_i(\alpha)-(\bar{I}_i+t(I_i-\bar{I}_i)))2w_i- 2w_i\left ( \frac{\partial \lambda_i(\alpha)}{\partial \alpha }  \right )^2 \right]} -  \left (2w_j\frac{\partial^2 \lambda_j(\alpha)}{\partial \alpha^2 }\right )\right ]\cdot \right. \\ 
& \left. \left (2w_i\frac{\partial \lambda_i(\alpha)}{\partial \alpha }\right )-\left (2w_i\frac{\partial^2 \lambda_i(\alpha)}{\partial \alpha^2 }\right ) \cdot \left (2w_j\frac{\partial \lambda_j(\alpha)}{\partial \alpha }\right ) \right ] \Biggr \rvert_{\alpha=\tau_{WLS}(\bar{I}^n+t(I^n-\bar{I}^n))},
\end{split}
\end{equation}
which concludes the result. 

\subsection{Proof of Proposition 2}
\label{sub_sec_prop_WLS}
\textit{Proof:} 
Using the function
$h(\alpha)=\sum_{i=1}^nw_i(\lambda_i(x_c)-\lambda_i(\alpha))^2$,
we need to show that the minimum is reached only at $\alpha=x_c$. From
this, we have that $h(\alpha)\geq 0$ and it achieves its minimum
at $x_c$.
To prove uniqueness, let us assume that there is another position
$x_c^*\neq x_c$ at which $h$ is zero. Then
\begin{eqnarray}
h(x_c^*)&=&\sum_{i=1}^nw_i(\lambda_i(x_c)-\lambda_i(x_c^*))^2=0\nonumber\\
&\Leftrightarrow& \lambda_i(x_c)=\lambda_i(x_c^*),~~ \forall i\in\{1,\ldots,n\}.
\end{eqnarray}
The last identity is not possible, because if we use a Gaussian PSF
there is at least one $i\in\{1,\ldots,n\}$ such that
$\lambda_i(x_c)\neq\lambda_i(x_c^*)$.
%

\subsection{Proof of Eq.~(\ref{eq_proof_th_wls_8})}
\label{proof_eq_proof_th_wls_8}
\textit{Proof:} Recall \cite[Eq.~(17)]{fessler1996} and considering $J_{WLS}(\alpha,I^n)$ as the cost function we have that
\begin{equation}
\begin{split}
\label{Fessler_id_wls}
\frac{\partial^2\tau_{WLS}}{\partial I_i \partial I_j} (\bar{I}^n+t(I^n-\bar{I}^n))&= \left [-\frac{\partial^2J_{WLS}(\alpha,\bar{I}^n+t(I^n-\bar{I}^n))}{\partial \alpha^2} \right ]^{-1}\left ( 
\left[\frac{\partial^3J_{WLS}(\alpha,\bar{I}^n+t(I^n-\bar{I}^n))}{\partial \alpha^3}\cdot \frac{\partial \tau_{WLS}(\bar{I}^n+t(I^n-\bar{I}^n))}{\partial I_j} \right . \right . +\\
 & \left. \left. \frac{\partial^3J_{WLS}(\alpha,\bar{I}^n+t(I^n-\bar{I}^n))}{\partial \alpha^2 \partial I_j} \right]\cdot \frac{\partial \tau_{WLS}(\bar{I}^n+t(I^n-\bar{I}^n))}{\partial I_i}\right.\\
 & \left . + \frac{\partial^3J_{WLS}(\alpha,\bar{I}^n+t(I^n-\bar{I}^n))}{\partial \alpha^2 \partial I_i} \cdot \frac{\partial \tau_{WLS}(\bar{I}^n+t(I^n-\bar{I}^n))}{\partial I_j} +\frac{\partial^3J_{WLS}(\alpha,\bar{I}^n+t(I^n-\bar{I}^n))}{\partial \alpha \partial I_i \partial I_j} \right )  \Biggr \rvert_{\alpha=\tau_{WLS}(\bar{I}^n+t(I^n-\bar{I}^n))},
 \end{split}
\end{equation}
where from the definition of $J_{ML}(\alpha, I^n)$ we have that
\begin{equation}
\label{wls_identity1}
\frac{\partial^2J_{WLS}(\alpha,\bar{I}^n+t(I^n-\bar{I}^n))}{\partial \alpha^2}=\sum_{i=1}^n \limits \frac{\partial^2 \lambda_i(\alpha)}{\partial \alpha ^2} \cdot (\lambda_i(\alpha)-(\bar{I}_i+t(I_i-\bar{I}_i)))2w_i+ 2w_i\left ( \frac{\partial \lambda_i(\alpha)}{\partial \alpha }  \right )^2,
\end{equation}
\begin{equation}
\label{wls_identity2}
\frac{\partial^3J_{WLS}(\alpha,\bar{I}^n+t(I^n-\bar{I}^n))}{\partial \alpha^3}=\sum_{i=1}^n \frac{\partial^3 \lambda_i(\alpha)}{\partial \alpha ^3} \cdot  (\lambda_i(\alpha)-(\bar{I}_i+t(I_i-\bar{I}_i)))2w_i +6w_i\frac{\partial^2 \lambda_i(\alpha)}{\partial \alpha ^2}\frac{\partial \lambda_i(\alpha)}{\partial \alpha },
\end{equation}
\begin{equation}
\label{wls_identity3}
\frac{\partial^3J_{WLS}(\alpha,\bar{I}^n+t(I^n-\bar{I}^n))}{\partial \alpha^2 \partial I_i}=-  \left (2w_i\frac{\partial^2 \lambda_i(\alpha)}{\partial \alpha^2 }\right ),
\end{equation}
\begin{equation}
\label{wls_identity4}
\frac{\partial^3J_{WLS}(\alpha,\bar{I}^n+t(I^n-\bar{I}^n))}{\partial \alpha \partial I_i \partial I_j}=0.
\end{equation}
Concerning $\frac{\partial \tau_{ML}(\bar{I}^n+t(I^n-\bar{I}^n))}{\partial I_i}$ it is just the $i$-th component of the gradient in Eq.~(\ref{eq_proof_th_wls_1}), then we use (\ref{eq_nabla20_astro_WLS}) and (\ref{eq_nabla11_astro_WLS})
\begin{equation}
\begin{split}
\label{wls_identity5}
 \frac{\partial \tau_{WLS}(\bar{I}^n+t(I^n-\bar{I}^n))}{\partial I_i}&= \frac{-\frac{\partial^2J_{WLS}(\alpha,\bar{I}^n+t(I^n-\bar{I}^n)))}{\partial \alpha \partial I_i}}{\frac{\partial^2J_{WLS}(\alpha,\bar{I}^n+t(I^n-\bar{I}^n))}{\partial \alpha^2}} \\
 &= \frac{\left (2w_i \frac{\partial \lambda_i(\alpha)}{\partial \alpha }\right )}{\sum_{i=1}^n \limits \frac{\partial^2 \lambda_i(\alpha)}{\partial \alpha ^2} \cdot (\lambda_i(\alpha)-(\bar{I}_i+t(I_i-\bar{I}_i)))2w_i+ 2w_i\left ( \frac{\partial \lambda_i(\alpha)}{\partial \alpha }  \right )^2}.
 \end{split}
\end{equation}
Finally, replacing (\ref{wls_identity1}), (\ref{wls_identity2}), (\ref{wls_identity3}), (\ref{wls_identity4}) and (\ref{wls_identity5}) in (\ref{Fessler_id_wls}), and evaluating in $\alpha=\tau_{WLS}(\bar{I}^n+t(I^n-\bar{I}^n))$ we obtain the desired result
\begin{equation}\label{eq_proof_th_wls_8b}
\begin{split}
\frac{\partial^2\tau_{WLS}}{\partial I_i \partial I_j} (\bar{I}^n+t(I^n-\bar{I}^n))&=
\frac{-1}{\left [ \sum_{i=1}^n \limits \frac{\partial^2 \lambda_i(\alpha)}{\partial \alpha ^2} \cdot (\lambda_i(\alpha)-(\bar{I}_i+t(I_i-\bar{I}_i)))2w_i+ 2w_i\left ( \frac{\partial \lambda_i(\alpha)}{\partial \alpha }  \right )^2 \right] ^2  }\cdot \\
& \left [ \left [ \left [\sum_{i=1}^n \frac{\partial^3 \lambda_i(\alpha)}{\partial \alpha ^3} \cdot  (\lambda_i(\alpha)-(\bar{I}_i+t(I_i-\bar{I}_i)))2w_i +6w_i\frac{\partial^2 \lambda_i(\alpha)}{\partial \alpha ^2}\frac{\partial \lambda_i(\alpha)}{\partial \alpha }  \right ] \cdot  \right.  \right. \\ 
&  \left. \left.\frac{\left (2w_j\frac{\partial \lambda_j(\alpha)}{\partial \alpha }\right )}{\left [ \sum_{i=1}^n \limits \frac{\partial^2 \lambda_i(\alpha)}{\partial \alpha ^2} \cdot (\lambda_i(\alpha)-(\bar{I}_i+t(I_i-\bar{I}_i)))2w_i- 2w_i\left ( \frac{\partial \lambda_i(\alpha)}{\partial \alpha }  \right )^2 \right]} -  \left (2w_j\frac{\partial^2 \lambda_j(\alpha)}{\partial \alpha^2 }\right )\right ]\cdot \right. \\ 
& \left. \left (2w_i\frac{\partial \lambda_i(\alpha)}{\partial \alpha }\right )-\left (2w_i\frac{\partial^2 \lambda_i(\alpha)}{\partial \alpha^2 }\right ) \cdot \left (2w_j\frac{\partial \lambda_j(\alpha)}{\partial \alpha }\right ) \right ] \Biggr \rvert_{\alpha=\tau_{WLS}(\bar{I}^n+t(I^n-\bar{I}^n))}.
\end{split}
\end{equation}

\section{Proof of Theorem 3}
\label{proof_th_ml}
\textit{Proof:} 
Again the proof and the derivation of $\sigma^2_{ML}(n)$,
$\beta_{ML}(n)$ and $\epsilon_{ML}(n)$ reduce to apply Theorem
1. First, we need to validate the assumption of
Theorem 1.  Beginning with the equality in
Eq.~(\ref{eq_proof_th_wls_1}),
it follows  that
\begin{eqnarray}
\nabla^{20}J_{ML}(\alpha,I^n)&=&\frac{\partial^2}{\partial \alpha^2}J_{ML}(\alpha,I^n)\nonumber\\
&=& -\sum_{i=1}^nI_i\frac{1}{\lambda_i^2(\alpha)}\left(\frac{\partial \lambda_i(\alpha)}{\partial \alpha}\right)^2+\sum_{i=1}^n\left(I_i\frac{1}{\lambda_i(\alpha)}-1\right)\frac{\partial^2\lambda_i(\alpha)}{\partial \alpha^2},\label{eq_nabla20_astro_ML}\\
\nabla^{11}J_{ML}(\alpha,I^n)&=& \left(\frac{\partial^2}{\partial \alpha \partial I_1}J_{ML}(\alpha,I^n),\ldots,\frac{\partial^2}{\partial \alpha\partial I_n}J_{ML}(\alpha,I^n)\right)^ T\nonumber\\
&=&\left( \frac{1}{\lambda_1(\alpha)}\frac{\partial \lambda_1(\alpha)}{\partial \alpha},\ldots, \frac{1}{\lambda_n(\alpha)}\frac{\partial \lambda_n(\alpha)}{\partial \alpha}\right)^ T.\label{eq_nabla11_astro_ML}
\end{eqnarray}
For evaluating these two expression at $\alpha=\tau_{ML}(\bar{I}^n)$
as required in (\ref{eq_proof_th_wls_1}), we use
that\footnote{Considering that $\bar{I}_i =E\{I_i\}=\lambda_i(x_c)$.}
\begin{equation}
\tau_{ML}(\bar{I}^n)=
\argmin_{\alpha\in\mathbb{R}}\sum_{i=1}^n-\lambda_i(x_c)
\ln(\lambda_i(\alpha))+\lambda_i(\alpha).
\label{tau_ML}
\end{equation}
Then we will use the following result
\begin{proposition}\label{prop_ML_A}
Under the assumption of a Gaussian PSF, $\tau_{ML}(\bar{I}^n)= x_c$.
\end{proposition}
Notice again, that this proposition is the second assumption used in Theorem
1. From this proposition, it follows that
\begin{eqnarray}
\nabla^{20}J(\tau_{ML}(\bar{I}^n),\bar{I}^n)
&=& -\sum_{i=1}^n\lambda_i(x_c)\frac{1}{\lambda_i^2(x_c)}\left.\left(\frac{\partial \lambda_i(\alpha)}{\partial \alpha}\right)^2\right|_{\alpha=x_c}+\sum_{i=1}^n\left(\frac{\lambda_i(x_c)}{\lambda_i(x_c)}-1\right)\left.\frac{\partial^2\lambda_i(\alpha)}{\partial \alpha^2}\right|_{\alpha=x_c},\nonumber\\
&=&-\sum_{i=1}^n\frac{1}{\lambda_i(x_c)}\left.\left(\frac{\partial \lambda_i(\alpha)}{\partial \alpha}\right)^2\right|_{\alpha=x_c},\label{eq_nabla20_astro2_ML}\\
\nabla^{11}J(\tau_{ML}(\bar{I}^n),\bar{I}^n)&=& \left.\left(\frac{\partial^2}{\partial \alpha \partial I_1}J(\alpha,I^n),\ldots,\frac{\partial^2}{\partial \alpha\partial I_n}J(\alpha,I)\right)^ T\right|_{\alpha=\tau_{ML}(\bar{I}^n)}\nonumber\\
&=&\left.\left( \frac{1}{\lambda_1(\alpha)}\frac{\partial \lambda_1(\alpha)}{\partial \alpha},\ldots, \frac{1}{\lambda_n(\alpha)}\frac{\partial \lambda_n(\alpha)}{\partial \alpha}\right)^ T\right|_{\alpha=x_c}.\label{eq_nabla11_astro2_ML}
\end{eqnarray}
Finally, we apply (\ref{eq_nabla20_astro2_ML}) and (\ref{eq_nabla11_astro2_ML}) in  (\ref{eq_nabla_theta}) to obtain that
\begin{eqnarray}
 \nabla \tau_{ML}(\bar{I}^n)\cdot (I^n-\bar{I}^n)&=& -[\nabla^{20}J(\tau_{ML}(\bar{I}^n),\bar{I}^n)]^{-1} [\nabla^{11}J(\tau_{ML}(\bar{I}^n),\bar{I}^n)](I^n-\bar{I}^n) \nonumber \\
 &=&\underbrace{-\left[\sum_{i=1}^n\frac{1}{\lambda_i(x_c)}\left. \left( \frac{\partial\lambda_i(\alpha)}{\partial \alpha}\right)^2\right|_{\alpha=x_c} \right]^{-1}}_{a}\cdot \sum_{j=1}^n\underbrace{\frac{1}{\lambda_j(x_c)}\left. \frac{\partial\lambda_j(\alpha)}{\partial \alpha} \right|_{\alpha=x_c} }_{b_j}\left ( I_j-\mathds{E}(I_j) \right ),
 \label{cov_bound_2}
\end{eqnarray}
that shows that the sufficient condition in (\ref{close_gradient}) of
Theorem 1 is satisfied. For computing the value $\sigma^2_{ML}(n)$ in (\ref{eq_proof_th_main_general_5}),  we have that
\begin{equation}\label{eq_cov_astro}
Cov\{I_i,I_j\}=\left\{
  \begin{array}{l l}
    Var\{I_i\}=\lambda_i(x_c), & \quad \text{if $i=j$},\\
    0 & \quad \sim.
  \end{array} \right.
\end{equation}
since the observations are independent and follow a Poisson distribution.
Then, replacing (\ref{eq_nabla20_astro2_ML}), (\ref{eq_nabla11_astro2_ML}) and (\ref{eq_cov_astro}) in 
(\ref{eq_proof_th_main_general_5}) we have that
\begin{eqnarray}
\sigma^2_{ML}(n)&=& \frac{1}{\sum_{i=1}^n\frac{1}{\lambda_i(x_c)}\left.\left(\frac{\partial \lambda_i(\alpha)}{\partial \alpha}\right)^2\right|_{\alpha=x_c}}, 
\end{eqnarray}
which resolves the identity in (\ref{eq_sec_main_astrometry_12}). Finally $\beta_{ML}(n)$ and $\epsilon_{ML}(n)$ comes from evaluating
(\ref{eq_proof_th_main_general_2}) and
(\ref{eq_proof_th_main_general_1}) in this ML context. For that we
only need to determine $\frac{\partial^2\tau_{ML}}{\partial I_i
  \partial I_j} (\bar{I}^n+t(I^n-\bar{I}^n))$.
Using \cite[Eq.~(17)]{fessler1996}, we can obtain the following
identity\footnote{The derivation of this result is presented in Appendix
  \ref{proof_ml_identity}.}
\begin{equation}\label{ml_identity}
\begin{split}
&\frac{\partial^2\tau_{ML}}{\partial I_i \partial I_j} (\bar{I}^n+t(I^n-\bar{I}^n))=
\frac{-1}{\left [ \sum_{i=1}^n \limits \frac{\partial^2 \lambda_i(\alpha)}{\partial \alpha ^2} \cdot \frac{\bar{I}_i+t(I_i-\bar{I}_i)}{\lambda_i(\alpha)}-\frac{\bar{I}_i+t(I_i-\bar{I}_i)}{\lambda_i^2(\alpha)}\cdot \left ( \frac{\partial \lambda_i(\alpha)}{\partial \alpha }  \right )^2 \right] ^2  }\cdot \\
& \left [ \left [ \left [\sum_{i=1}^n \limits \frac{\partial^3 \lambda_i(\alpha)}{\partial \alpha ^3} \cdot  \frac{\bar{I}_i+t(I_i-\bar{I}_i)}{\lambda_i(\alpha)} - 3\frac{\partial^2 \lambda_i(\alpha)}{\partial \alpha ^2}\frac{\partial \lambda_i(\alpha)}{\partial \alpha }  \frac{\bar{I}_i+t(I_i-\bar{I}_i)}{\lambda_i^2(\alpha)}+2\frac{\bar{I}_i+t(I_i-\bar{I}_i)}{\lambda_i^3(\alpha)} \left ( \frac{\partial \lambda_i(\alpha)}{\partial \alpha }  \right )^3   \right ] \cdot  \right.  \right. \\ 
&  \left. \left.\frac{\left (-\frac{1}{\lambda_j(\alpha)} \frac{\partial \lambda_j(\alpha)}{\partial \alpha }\right )}{\left [ \sum_{i=1}^n \frac{\partial^2 \lambda_i(\alpha)}{\partial \alpha ^2} \cdot \frac{\bar{I}_i+t(I_i-\bar{I}_i)}{\lambda_i(\alpha)}-\frac{\bar{I}_i+t(I_i-\bar{I}_i)}{\lambda_i^2(\alpha)}\cdot \left ( \frac{\partial \lambda_i(\alpha)}{\partial \alpha }  \right )^2 \right]} +  \frac{\partial^2 \lambda_j(\alpha)}{\partial \alpha ^2}\frac{1}{\lambda_j(\alpha)}- \frac{1}{\lambda_j(\alpha)}\left ( \frac{\partial \lambda_j(\alpha)}{\partial \alpha }  \right )^2 \right ]\cdot \right. \\ 
& \left. \left (-\frac{1}{\lambda_i(\alpha)} \frac{\partial \lambda_i(\alpha)}{\partial \alpha }\right )+ \left (\frac{\partial^2 \lambda_i(\alpha)}{\partial \alpha ^2}\frac{1}{\lambda_i(\alpha)}- \frac{1}{\lambda_i(\alpha)}\left ( \frac{\partial \lambda_i(\alpha)}{\partial \alpha }  \right )^2 \right) \cdot \left (-\frac{1}{\lambda_j(\alpha)} \frac{\partial \lambda_j(\alpha)}{\partial \alpha }\right ) \right ] \Biggr \rvert_{\alpha=\tau_{ML}(\bar{I}^n+t(I^n-\bar{I}^n))},
\end{split}
\end{equation}
which concludes the result.

\subsection{Proof of Proposition 3}
\textit{Proof:} 
Let us consider the function $g_n:\R_+^n \rightarrow \R$ given by
\begin{equation}
g_n(y_1,...,y_n)_{\lambda_i^n}= \sum_{i=1}^n -\lambda_i \ln(y_i)+y_i.
\end{equation}
We note that
\begin{equation}
\min_{y_1^n \in \R_+^n}g_n(y_1,...,y_n)_{\lambda_i^n}=\sum_{i=1}^n\min_{y_i \in \R_+}g_1(y_i)_{\lambda_i},
\end{equation}
where applying first order condition
$y_i =\lambda_i$, $\forall i \in \{1,...,n \}$.
Returning to our problem in (\ref{tau_ML}) where $\lambda_i=\bar{I}_i=\lambda_i(x_c)$ and $y_i=\lambda_i(\alpha)$,  
it is clear, considering the Gaussian profile in PSF,  that
\begin{equation}
\lambda_i(\alpha) =\lambda_i(x_c) \ \ \ \ \forall i \in \{1,...,n \} \text{ if } \alpha=x_c,
\end{equation}
which concludes the result.

\subsection{Proof of Eq.~(\ref{ml_identity})}
\label{proof_ml_identity}
\textit{Proof:} Recall \cite[Eq.~(17)]{fessler1996} and considering $J_{ML}(\alpha,I^n)$ as the cost function we have that:
\begin{equation}
\begin{split}
\label{Fessler_id_ml}
\frac{\partial^2\tau_{ML}}{\partial I_i \partial I_j} (\bar{I}^n+t(I^n-\bar{I}^n))&= \left [-\frac{\partial^2J_{ML}(\alpha,\bar{I}^n+t(I^n-\bar{I}^n))}{\partial \alpha^2} \right ]^{-1}\left ( 
\left[\frac{\partial^3J_{ML}(\alpha,\bar{I}^n+t(I^n-\bar{I}^n))}{\partial \alpha^3}\cdot \frac{\partial \tau_{ML}(\bar{I}^n+t(I^n-\bar{I}^n))}{\partial I_j} \right . \right . +\\
 & \left. \left. \frac{\partial^3J_{ML}(\alpha,\bar{I}^n+t(I^n-\bar{I}^n))}{\partial \alpha^2 \partial I_j} \right]\cdot \frac{\partial \tau_{ML}(\bar{I}^n+t(I^n-\bar{I}^n))}{\partial I_i}\right.\\
 & \left . + \frac{\partial^3J_{ML}(\alpha,\bar{I}^n+t(I^n-\bar{I}^n))}{\partial \alpha^2 \partial I_i} \cdot \frac{\partial \tau_{ML}(\bar{I}^n+t(I^n-\bar{I}^n))}{\partial I_j} +\frac{\partial^3J_{ML}(\alpha,\bar{I}^n+t(I^n-\bar{I}^n))}{\partial \alpha \partial I_i \partial I_j} \right )  \Biggr \rvert_{\alpha=\tau_{ML}(\bar{I}^n+t(I^n-\bar{I}^n))},
 \end{split}
\end{equation}
where from the definition of $J_{ML}(\alpha, I^n)$ we have that
\begin{equation}
\label{ml_identity1}
\frac{\partial^2J_{ML}(\alpha,\bar{I}^n+t(I^n-\bar{I}^n))}{\partial \alpha^2}=\sum_{i=1}^n \limits \frac{\partial^2 \lambda_i(\alpha)}{\partial \alpha ^2} \cdot \frac{\bar{I}_i+t(I_i-\bar{I}_i)}{\lambda_i(\alpha)}-\frac{\bar{I}_i+t(I_i-\bar{I}_i)}{\lambda_i^2(\alpha)}\cdot \left ( \frac{\partial \lambda_i(\alpha)}{\partial \alpha }  \right )^2,
\end{equation}
\begin{equation}
\label{ml_identity2}
\frac{\partial^3J_{ML}(\alpha,\bar{I}^n+t(I^n-\bar{I}^n))}{\partial \alpha^3}=\sum_{i=1}^n \limits \frac{\partial^3 \lambda_i(\alpha)}{\partial \alpha ^3} \cdot  \frac{\bar{I}_i+t(I_i-\bar{I}_i)}{\lambda_i(\alpha)} - 3\frac{\partial^2 \lambda_i(\alpha)}{\partial \alpha ^2}\frac{\partial \lambda_i(\alpha)}{\partial \alpha }  \frac{\bar{I}_i+t(I_i-\bar{I}_i)}{\lambda_i^2(\alpha)}+2\frac{\bar{I}_i+t(I_i-\bar{I}_i)}{\lambda_i^3(\alpha)} \left ( \frac{\partial \lambda_i(\alpha)}{\partial \alpha }  \right )^3,
\end{equation}
\begin{equation}
\label{ml_identity3}
\frac{\partial^3J_{ML}(\alpha,\bar{I}^n+t(I^n-\bar{I}^n))}{\partial \alpha^2 \partial I_i}=\frac{\partial^2 \lambda_i(\alpha)}{\partial \alpha ^2} \cdot \frac{1}{\lambda_i(\alpha)}-\frac{1}{\lambda_i^2(\alpha)}\cdot \left ( \frac{\partial \lambda_i(\alpha)}{\partial \alpha }  \right )^2,
\end{equation}
\begin{equation}
\label{ml_identity4}
\frac{\partial^3J_{ML}(\alpha,\bar{I}^n+t(I^n-\bar{I}^n))}{\partial \alpha \partial I_i \partial I_j}=0.
\end{equation}
Concerning $\frac{\partial
  \tau_{ML}(\bar{I}^n+t(I^n-\bar{I}^n))}{\partial I_i}$ it is just the
$i$-th component of the gradient in Eq.~(\ref{eq_proof_th_wls_1}),
then we use (\ref{eq_nabla20_astro_ML}) and
(\ref{eq_nabla11_astro_ML})
\begin{equation}
\begin{split}
\label{ml_identity5}
 \frac{\partial \tau_{ML}(\bar{I}^n+t(I^n-\bar{I}^n))}{\partial I_i}&= \frac{-\frac{\partial^2J_{ML}(\alpha,\bar{I}^n+t(I^n-\bar{I}^n)))}{\partial \alpha \partial I_i}}{\frac{\partial^2J_{ML}(\alpha,\bar{I}^n+t(I^n-\bar{I}^n))}{\partial \alpha^2}} \\
 &= \frac{\left (-\frac{1}{\lambda_i(\alpha)} \frac{\partial \lambda_i(\alpha)}{\partial \alpha }\right )}{\sum_{i=1}^n \limits \frac{\partial^2 \lambda_i(\alpha)}{\partial \alpha ^2} \cdot \frac{\bar{I}_i+t(I_i-\bar{I}_i)}{\lambda_i(\alpha)}-\frac{\bar{I}_i+t(I_i-\bar{I}_i)}{\lambda_i^2(\alpha)}\cdot \left ( \frac{\partial \lambda_i(\alpha)}{\partial \alpha }  \right )^2}.
 \end{split}
\end{equation}
Finally, replacing (\ref{ml_identity1}), (\ref{ml_identity2}), (\ref{ml_identity3}), (\ref{ml_identity4}) and (\ref{ml_identity5}) in (\ref{Fessler_id_ml}), and evaluating in $\alpha=\tau_{ML}(\bar{I}^n+t(I^n-\bar{I}^n))$ we obtain the desired result

\begin{equation}
\begin{split}
&\frac{\partial^2\tau_{ML}}{\partial I_i \partial I_j} (\bar{I}^n+t(I^n-\bar{I}^n))=
\frac{-1}{\left [ \sum_{i=1}^n \limits \frac{\partial^2 \lambda_i(\alpha)}{\partial \alpha ^2} \cdot \frac{\bar{I}_i+t(I_i-\bar{I}_i)}{\lambda_i(\alpha)}-\frac{\bar{I}_i+t(I_i-\bar{I}_i)}{\lambda_i^2(\alpha)}\cdot \left ( \frac{\partial \lambda_i(\alpha)}{\partial \alpha }  \right )^2 \right] ^2  }\cdot \\
& \left [ \left [ \left [\sum_{i=1}^n \limits \frac{\partial^3 \lambda_i(\alpha)}{\partial \alpha ^3} \cdot  \frac{\bar{I}_i+t(I_i-\bar{I}_i)}{\lambda_i(\alpha)} - 3\frac{\partial^2 \lambda_i(\alpha)}{\partial \alpha ^2}\frac{\partial \lambda_i(\alpha)}{\partial \alpha }  \frac{\bar{I}_i+t(I_i-\bar{I}_i)}{\lambda_i^2(\alpha)}+2\frac{\bar{I}_i+t(I_i-\bar{I}_i)}{\lambda_i^3(\alpha)} \left ( \frac{\partial \lambda_i(\alpha)}{\partial \alpha }  \right )^3   \right ] \cdot  \right.  \right. \\ 
&  \left. \left.\frac{\left (-\frac{1}{\lambda_j(\alpha)} \frac{\partial \lambda_j(\alpha)}{\partial \alpha }\right )}{\left [ \sum_{i=1}^n \frac{\partial^2 \lambda_i(\alpha)}{\partial \alpha ^2} \cdot \frac{\bar{I}_i+t(I_i-\bar{I}_i)}{\lambda_i(\alpha)}-\frac{\bar{I}_i+t(I_i-\bar{I}_i)}{\lambda_i^2(\alpha)}\cdot \left ( \frac{\partial \lambda_i(\alpha)}{\partial \alpha }  \right )^2 \right]} +  \frac{\partial^2 \lambda_j(\alpha)}{\partial \alpha ^2}\frac{1}{\lambda_j(\alpha)}- \frac{1}{\lambda_j(\alpha)}\left ( \frac{\partial \lambda_j(\alpha)}{\partial \alpha }  \right )^2 \right ]\cdot \right. \\ 
& \left. \left (-\frac{1}{\lambda_i(\alpha)} \frac{\partial \lambda_i(\alpha)}{\partial \alpha }\right )+ \left (\frac{\partial^2 \lambda_i(\alpha)}{\partial \alpha ^2}\frac{1}{\lambda_i(\alpha)}- \frac{1}{\lambda_i(\alpha)}\left ( \frac{\partial \lambda_i(\alpha)}{\partial \alpha }  \right )^2 \right) \cdot \left (-\frac{1}{\lambda_j(\alpha)} \frac{\partial \lambda_j(\alpha)}{\partial \alpha }\right ) \right ] \Biggr \rvert_{\alpha=\tau_{ML}(\bar{I}^n+t(I^n-\bar{I}^n))}.
\end{split}
\end{equation}

\end{appendix}

\bibliographystyle{aa}
\bibliography{ref}

\end{document}